\documentclass[12pt]{JHEP}
\usepackage{graphicx}
\usepackage{bm}
\usepackage{amsmath,epsfig}
\usepackage{amssymb,amsfonts}
\usepackage{latexsym}
\usepackage{epsfig}
\newcommand{\refeq}[1]{(\ref{#1})}
\newcommand{\intd}{\mathrm{d}}

\newcommand {\non}{\nonumber}

\def\hri#1#2{\href{http://arxiv.org/abs/#1}{[ArXiv:#1]#2}}
\def\hre#1#2{\href{http://arxiv.org/abs/#1/#2}{[ArXiv:#1/#2]}}

\def\be{\begin{equation}}
\def\ee{\end{equation}}
\def\bea{\begin{eqnarray}}
\def\eea{\end{eqnarray}}

\newcommand\fverb{\setbox\pippobox=\hbox\bgroup\verb}
\newcommand\fverbdo{\egroup\medskip\noindent%
                        \fbox{\unhbox\pippobox}\ }
\newcommand\fverbit{\egroup\item[\fbox{\unhbox\pippobox}]}

\newcommand{\bear}{\begin{eqnarray}}
\newcommand{\eear}{\end{eqnarray}}
\newcommand{\de}{\partial}
\newcommand{\<}{\langle}
\renewcommand{\>}{\rangle}
\newbox\pippobox

\def\lab{\label}
\def\6{\partial}
\def\f{\Phi}
\def\a{\alpha}
\def\nn{\nonumber}

\def\le{\left}
\def\ri{\right}

\def\pa{\partial}

\def\e{\epsilon}
\def\m{\mu}

\def\sp{\;\;\;,\;\;\;}

\def\sq
\def\a{\alpha}
\def\b{\beta}
\def\l{\lambda}

\def\tr{{\rm Tr}}

\def\C0{{\bf C_0}}
\def\Y0{{\bf Y_0}}
\def\G0{{\bf G_0}}

\title{Improved Holographic Yang-Mills at Finite Temperature: Comparison with Data}

\author{U. G{\"u}rsoy$^1$,
\href{http://hep.physics.uoc.gr/~kiritsis/}{E. Kiritsis}$^{2}$,
L. Mazzanti$^{3}$,
 F. Nitti$^{4}$\\
$^1$\href{http://www1.phys.uu.nl/wwwitf}{Institute for Theoretical Physics, Utrecht University;
Leuvenlaan 4, 3584 CE Utrecht, The Netherlands.}\\
~\\
$^2$\href{http://hep.physics.uoc.gr/}{Department of Physics, University of Crete
71003 Heraklion, Greece}\\
~\\
$^3$\href{http://cpht.polytechnique.fr/cpht/cordes/}{CPHT, Ecole Polytechnique, CNRS,
 91128, Palaiseau, France}\\
 (UMR du CNRS 7644).\\
~\\
$^4$\href{http://www.apc.univ-paris7.fr}{APC, Universit\'e Paris 7, \\ B\^atiment Condorcet, F-75205, Paris Cedex 13, France}}

\preprint{CPHT-RR012.0309}      

\abstract{The semi-phenomenological improved holographic model for QCD is
confronted with data of the pure glue, large-$N_c$ gauge theory.
After fitting two phenomenological parameters in the potential, the model can
reproduce in detail all thermodynamic functions at finite temperature.
It also reproduces in detail all known spin-0 and spin-2 glueball observables
at zero temperature and predicts the rest of the $0^{++}$ and $2^{++}$ towers.
A similar two parameter fit in the CP-odd sector postdicts the correct second $0^{+-}$ glueball mass,
and predicts the rest of the $0^{+-}$ tower.}
\begin{document}

\maketitle

\section{Introduction}

The experimental efforts at RHIC, \cite{rhic} have provided a novel window in the physics of the strong interactions.
The consensus on the existing data is that shortly after the collision, a ball of quark-gluon plasma (QGP) forms that is at thermal equilibrium,
and  subsequently expands until its temperature falls below the QCD transition (or crossover) where it finally hadronizes.
Relativistic hydrodynamics describes very well the QGP  \cite{lr}, with a shear-viscosity to entropy density ratio close to that
of ${\cal N}=4$ SYM, \cite{pss}.
The QGP is at strong coupling, and it necessitates a treatment beyond perturbative QCD approaches, \cite{review}.
Moreover, although the shear viscosity from  ${\cal N}=4$ seems to be close to that ``measured'' by
 experiment, lattice data indicate that in the relevant
RHIC range  $1\leq {T\over T_c}\leq 3$ the QGP
seems not to be a fully  conformal fluid.
 Therefore the bulk viscosity may play a role near the phase transition
\cite{kkt,m}.  The lattice techniques have been successfully used
to study the thermal behavior of QCD, however they are not easily
extended to the computation of hydrodynamic quantities. They can
be used however, together with parametrizations of correlators in
order to pin down parameters \cite{m}. On the other hand,
approaches based on holography have the potential to address
directly the real-time strong coupling physics relevant for
experiment.

In the bottom-up holographic model of AdS/QCD \cite{adsqcd1}, the bulk viscosity is zero as conformal invariance is essentially not broken
(the stress tensor is traceless).
In the soft-wall model \cite{soft}, no reliable calculation can be done for glue correlators and therefore
transport coefficients are ill-defined. Similar
remarks hold for other phenomenologically interesting observables as the
drag force and the jet quenching parameter \cite{her,lrw,gub,tea}.

Top-down holographic models of QCD displaying all relevant features of the theory have been difficult to obtain.
Bottom-up models based on AdS slices \cite{ps} have given some insights mostly in the meson sector, \cite{adsqcd1}
but necessarily lack many important holographic features of QCD.
A hybrid approach has been advocated \cite{ihqcd,diss} combining features of bottom-up and top-down models.
Such an approach is essentially a five-dimensional dilaton-gravity system with a non-trivial dilaton potential.
Flavor can be eventually added in the form of $N_f$ space-time filling
$D4-\overline{D4}$ brane pairs, supporting $U(N_f)_L\times U(N_f)_R$ gauge fields and a bi-fundamental
scalar \cite{ckp}.
The UV asymptotics of the potential are fixed by QCD perturbation theory, while the IR asymptotics of the potential can be fixed by confinement and linear glueball asymptotics.
An analysis of the finite temperature behavior \cite{GKMN1,GKMN2} has shown that the phase structure is exactly what one would expect from YM.
A potential with a single free parameter tuned to match
 the zero temperature glueball spectrum was able to agree with the thermodynamic behavior of glue
to a good degree, \cite{GKMN1}. Similar results, but with somewhat different potentials were also obtained in \cite{gubser,dew}

In \cite{GKMN1,GKMN2} it was shown that Einstein-dilaton gravity with a strictly monotonic dilaton potential that grows sufficiently
fast, generically shares the same phase structure and
thermodynamics of finite-temperature pure Yang-Mills theory at large $N_c$.
There is a deconfinement phase transition (dual to a Hawking-Page phase transition between a black-hole and thermal gas
background on the gravity side), which is generically first order.
 The latent heat scales as $N_c^2$. In the deconfined gluon-plasma phase,
the free energy slowly approaches that of a free gluon gas at high temperature,
and the speed of sound starts from a small value at $T_c$ and approaches the conformal value $c_s^2=1/3$ as the temperature increases. The deviation from conformal invariance is strongest at $T_c$, and
is signaled by the presence of a non-trivial gluon condensate, which on the gravity side emerges as
a deviation of the scalar solution that behaves asymptotically as $r^4$ close to the UV boundary. In the CP-violating
sector, the topological vacuum density $\tr F\tilde{F}$ has zero expectation value in the deconfined phase,
in agreement with lattice results \cite{ltheta} and large-$N_c$ expectations.

The analysis performed in \cite{GKMN2} was completely general and did not rely on any specific form of the dilaton potential $V(\l)$.
In this paper  we present instead a detailed  analysis of an explicit model, whose thermodynamics matches {\em quantitatively} the thermodynamics of pure Yang-Mills theory.
The (dimensionless) free energy, entropy density, latent heat and speed of sound, obtained on the gravity side by numerical integration of the 5D
field equations, are compared with the corresponding quantities, calculated on the lattice  for pure Yang-Mills at finite-$T$, resulting in excellent
agreement, for the temperature range that is accessible by lattice techniques.
The same  model also shows a good agreement with the lattice calculation of glueball mass ratios at zero temperature,
and  we find that the  value of the deconfining
critical temperature (in units of the lowest glueball mass)  is also in good agreement with the lattice results.

In short, the model we present gives a good phenomenological holographic description of
most static  properties\footnote{There are very few observables also that are not in agreement with YM.
 They are discussed in detail in \cite{diss}.} (spectrum and equilibrium thermodynamics)  of
large-$N_c$ pure Yang-Mills, as computed on the lattice, for energies up to several  times $T_c$.  Thus it constitutes a good starting point for the
computation of dynamical observables in a {\em realistic} holographic dual to QCD (as opposed to e.g. ${\cal N}=4$ SYM), such as transport
coefficients and other hydrodynamic properties that are not easily accessible by lattice techniques, at energies and temperatures relevant for
relativistic heavy-ion collision experiments. We will report on such a calculation in the near future.

The vacuum solution in this model is described in terms of two basic bulk fields, the metric and the dilaton.
These are not the only bulk fields however, as the bulk theory is expected to have an a priori infinite number
of fields, dual to all possible YM operators.
In particular we know from the string theory side that there are a few other low mass fields, namely the RR axion (dual to the QCD $\theta$-angle)
the NSNS and RR two forms $B_2$ and $C_2$ as well as other higher-level fields. With the exception of the RR axion, such fields are dual
to higher-dimension and/or higher-spin operators of YM.
Again, with the exception of the RR axion, they are not expected to play an important
 role into the structure of the vacuum and this is why we neglect them when we solve the equations
of motion. However, they are going to generate several new towers of glueball states
 beyond those that we discuss in this paper (namely the $0^{++}$
glueballs associated to dilaton fluctuations, $2^{++}$
glueballs associated to graviton  fluctuations and $0^{-+}$
glueballs associated to RR axion  fluctuations). Such fields can be included in
 the effective action and the associated glueball spectra calculated.
Since we do not know the detailed structure of the associated string theory,
their effective action will depend on more semi-phenomenological functions like $Z(\lambda)$ in (\ref{axionaction}).
These functions can again be determined in a way similar to $Z(\lambda)$.
In particular including the $B_2$ and $C_2$ field will provide $1^{+-}$ glueballs among others.
Fields with spin greater than 2 are necessarily stringy in origin.
We will not deal further with extra fields, like $B_2$ and $C_2$ and other as they are not particularly relevant for the purposes of
this model, namely the study of finite temperature physics in the deconfined case. We will only consider the axion,
as its physics is related to the CP-odd sector of YM with an obvious phenomenological importance.

It is well documented that string theory duals of YM must have strong curvatures in the UV regime. This has been explained in detail in \cite{diss}
where it was also argued, that although the asymptotic AdS boundary geometry is due to the curvature non-linearities of the associated string theory,
the inwards geometry is perturbative around AdS, with logarithmic corrections, generating the YM perturbation theory.
The present model is constructed so that it takes the asymptotic AdS geometry for granted, by introducing the associated vacuum energy by hand,
and simulates the perturbative YM expansion by an appropriate dilaton potential.
In the IR, we do not expect strong curvatures in the string frame, and indeed the preferred backgrounds have this property.
In this sense the model contains in itself the relevant expected effects that should arise from strong curvatures in all regimes.
These issues have been explained in \cite{ihqcd} and in more detail in \cite{diss}.

The paper is organized a follows. In Section \ref{model} we review the general features of the 5D model and its thermodynamics.
In Section \ref{scheme} we discuss the issue of the possible sources of scheme dependence in the relation between  the 5D dilaton
field and the 4D Yang-Mills coupling. In Section \ref{parameters} we fix the form of the dilaton potential and we  discuss in detail
the independent parameters appearing in the model and the role they play in the dynamics.  Section \ref{matching}  contains the main results: the comparison between our model and
the lattice data for the thermodynamics of the deconfined phase, and for the spectrum of the confined phase. Finally, In Section \ref{conclusions} we provide some general conclusions and discuss future directions.
Appendix \ref{numerics} contains a description of
the strategy we employ to numerically integrate Einstein's equations in order to find black hole solutions with different
temperatures but same UV asymptotics.

\section{The 5D model} \label{model}

The holographic dual of large $N_c$ Yang Mills theory,  proposed in \cite{ihqcd},  is
based on a five-dimensional Einstein-dilaton model, with the action:
\begin{equation}
S_5=-M^3_pN_c^2\int d^5x\sqrt{g}
\left[R-{4\over 3}(\partial\Phi)^2+V(\Phi) \right]+2M^3_pN_c^2\int_{\partial M}d^4x \sqrt{h}~K.
 \label{a1}\end{equation}
Here, $M_p$ is the  five-dimensional Planck scale and $N_c$ is the number of colors.
The last term is the Gibbons-Hawking term, with $K$ being the extrinsic curvature
of the boundary. The effective five-dimensional Newton constant
is $G_5 = 1/(16\pi M_p^3 N_c^2)$, and it is small in the large-$N_c$ limit.

Of the 5D coordinates $\{x_i, r\}_{i=0\ldots 3}$, $x_i$ are identified with the
4D space-time coordinates, whereas  the  radial coordinate $r$ corresponds to the 4D RG scale.
We identify $\l\equiv e^\Phi$ with the  running 't Hooft  coupling $\l_t\equiv N_cg_{YM}^2$,
up to an {\it a priori} unknown multiplicative factor\footnote{ This relation is well motivated
in the UV, although it may be modified at strong coupling (see Section \ref{scheme}). The
quantities we will calculate do not depend on the explicit relation between $\l$ and $\l_t$.
}, $\l = \kappa \l_t$.

The dynamics is encoded in the dilaton potential\footnote{With a slight abuse of notation we will denote $V(\l)$  the
function $V(\Phi)$ expressed as a function of  $\l\equiv e^\Phi$.},  $V(\l)$.
The small-$\l$ and large-$\l$ asymptotics of $V(\l)$ determine the solution in the UV
and  the IR of the geometry
respectively. For a detailed but concise description of the UV and IR  properties of the solutions the reader
is referred to Section 2 of \cite{GKMN2}. Here we will only mention the most relevant information:
\begin{enumerate}
\item For small $\l$,   $V(\l)$  is required to have a power-law expansion of the form:
\be \label{UVexp}
V(\l) \sim {12\over \ell^2}(1+ v_0 \l + v_1 \l^2 +\ldots), \qquad \l\to 0 \;.
\ee
The value at $\l=0$ is constrained to be finite and positive, and sets the UV $AdS$ scale $\ell$.
 The coefficients
of the other terms in the expansion fix the  $\beta$-function coefficients for the
running coupling $\l(E)$. If we  identify the energy scale with the metric scale factor in the Einstein frame, as
in  \cite{ihqcd}, we have:
\bea\label{betafunc}
&&\beta(\l) \equiv {d \lambda \over d\log E} = -b_0\l^2 -b_1 \l^3 +\ldots\nn\\
&& b_0 = {9\over 8} v_0, \quad \; \; b_1 = \frac94 v_1 - \frac{207}{256}v_0^2 \;.
\eea
\item For large $\l$,   confinement and the absence of bad singularities\footnote{We call ``bad  singularities'' those that
do not have  a well defined spectral problem for the fluctuations
without imposing extra boundary conditions.} require:
\be\label{IRexp}
 V(\l) \sim \l^{2Q}(\log \l)^P \quad \l\to \infty, \quad \left\{ \begin{array}{l} 2/3 < Q < 2\sqrt{2}/3, \quad P\; {\rm arbitrary}\\ Q = 2/3, \quad P\geq 0 \end{array} \right. .
\ee
In particular, the values $Q=2/3, P=1/2$ reproduce a linear glueball spectrum, $m_n^2\sim n$, besides confinement.
We will restrict ourselves to this case in what follows.
\end{enumerate}

In the large $N_c$ limit,
the  canonical ensemble partition function of the model just described, can be
approximated by a sum over saddle points, each given by a classical solution of the Einstein-dilaton
field equations:
\be
{\cal Z}(\beta) \simeq e^{-{\cal S}_1(\beta)}  +   e^{-{\cal S}_2(\beta)} + \ldots
\ee
where $S_i$ are the euclidean actions evaluated on each  classical solution with a fixed
 temperature $T=1/\beta$, i.e. with euclidean time compactified on a circle of length $\beta$.
There are two possible types of Euclidean solutions which preserve 3-dimensional rotational invariance.
In conformal coordinates these are:
\begin{enumerate}
\item {\bf Thermal gas solution,}
\be\label{thermal}
ds^2 = b^2_o(r)\left(dr^2 + dt^2 +  dx_mdx^m\right), \qquad \Phi = \Phi_o(r),
\ee
with $r\in (0, \infty)$ for the values of $P$ and $Q$ we are using;
\item {\bf Black hole solutions,}
\be
ds^2=b(r)^2\left[{dr^2\over f(r)}+f(r)dt^2+dx_mdx^m\right], \qquad \Phi = \Phi(r),
 \label{a7}\ee
with  $r\in (0,r_h)$,  such that $f(0)=1$, and $f(r_h)=0$.
\end{enumerate}
In both cases Euclidean time is periodic with period $\b_o$ and $\b$ respectively for the thermal gas and black-hole solution, 
and 3-space is
taken to be a torus with volume $V_{3o}$ and $V_3$ respectively,
so that the black hole mass and entropy are finite\footnote{The periods
and 3-space volumes of the thermal gas solution are related to the black-hole solution values by requiring that the geometry of the two solutions are
the same on the (regulated) boundary. See \cite{GKMN2} for details.}.

The black holes are dual to a deconfined phase, since the string
tension vanishes at the horizon, and the Polyakov loop has
non-vanishing expectation value (\cite{D4,sonnenschein}). On the
other hand, the thermal  gas background is confining.

The thermodynamics of the deconfined phase is dual to the  5D
black hole thermodynamics. The  free energy, defined as
 \be\label{first law}
{\cal F} = E - T S,
\ee
is identified with the  black hole on-shell
action; as usual, the energy $E$ and entropy $S$ are identified  with the black hole mass, and one
fourth of the horizon area in Planck units,  respectively.



The thermal gas and black hole solutions with the same temperature differ at $O(r^4)$:
\be
b(r) = b_o(r)\left[1 + \,{\cal G}\, {r^4\over \ell^3} +\ldots\right],  \qquad f(r) = 1 -{C\over 4} {r^4\over \ell^3} + \ldots \qquad r \to 0,
\label{b-bo}
\ee where ${\cal G}$ and $C$ are constants with units of energy.
As shown in \cite{GKMN2} they  are related to enthalpy $TS$ and
the gluon condensate $\<\tr F^2\>$ : \be\label{CG} C = {T S \over
M_p^3 N_c^2 V_3} , \qquad \qquad {\cal G} ={22 \over 3 (4\pi)^{2}}
{\langle \tr~F^2 \rangle_T - \langle \tr~F^2 \rangle_o \over240
M_p^3 N_c^2}. \ee Although they appear as coefficients in the UV
expansion, $C$ and ${\cal G}$ are determined by regularity at the
black hole horizon. For $T$ and $S$ the relation is the usual one,
\be\label{TS}
 T = - {\dot{f}(r_h) \over 4\pi}, \qquad S  = {Area \over 4 G_5} = 4\pi\, (M_p^3 N_c^2 V_3) \, b^3(r_h).
\ee
For ${\cal G}$ the relation with the horizon quantities is more complicated and
cannot be put in a simple analytic form. However, as discussed in \cite{GKMN2}, for each temperature
there exist only specific values of ${\cal G}$ (each corresponding to a different black hole)
such that the horizon is regular.

At any given temperature there can be one or more solutions: the thermal gas
is always present, and there can be different black holes with the same temperature. The solution  that dominates
the partition function at a certain $T$ is the one with smallest free energy. The free energy difference between the black hole  and
thermal gas was calculated in \cite{GKMN2}  to be:
\be\label{F}
\frac{\cal F}{M_p^3 N_c^2 V_3}={{\cal F}_{BH} - {\cal F}_{th}\over M_p^3 N_c^2 V_3}  = 15 {\cal G} - {C\over 4}.
\ee
For a dilaton potential corresponding to a confining theory, like the one we will assume,
the phase structure is the following \cite{GKMN2}:
\begin{enumerate}
\item There exists a minimum temperature $T_{min}$ below which the only solution is the thermal gas.
\item Two branches of black holes (``big'' and ``small'')  appear for $T\geq T_{min}$, but the ensemble
is still dominated by the confined phase up to a temperature $T_c > T_{min}$
\item At $T=T_c$ there is a first order phase transition to the black hole phase. The system remains
in the black hole (deconfined) phase for all $T>T_c$.
\end{enumerate}
In principle there could be more than two black hole branches, but this
will not happen with the specific potential we will use.

\section{Scheme dependence}\label{scheme}

There are several sources of scheme dependence in any attempt to solve a QFT.
Different parametrizations of the coupling constant (here $\l$) give different descriptions.
However,  physical statements must be invariant under such a change.
In our case,  reparametrizations of the coupling constant are equivalent to radial diffeomorphisms as we could use
$\l$ as the radial coordinate.

In the holographic context,  scheme dependence related to coupling redefinitions translates into field redefinitions for the bulk fields.
As the bulk theory is on-shell, all on-shell observables (that are evaluated at the single boundary of space-time) are independent
of the field redefinitions showing that scheme-independence is expected.
Invariance under radial reparametrizations of scalar bulk invariants is equivalent to RG invariance.
Because of renormalization effects, the boundary is typically shifted and in this case field redefinitions must be combined
with appropriate radial diffeomorphisms that amount to RG-transformations.

Another source of scheme dependence in our setup comes from the choice of
 the energy function. Again we may also consider this as a radial coordinate and therefore
it is subject to coordinate transformations.
A relation between $\l$ and $E$ is the $\beta$ function,
\be
{d\l\over d\log E}=\beta(\l).
\ee
$\beta$ by definition transforms as a vector under $\l$ reparametrizations
and as a form under $E$ reparametrizations.
$\beta(\l)$ can therefore be thought of as a vector field implementing the change of coordinates from $\lambda$ to $E$ and vice-versa.

Physical quantities should be independent of scheme. They are
quantities that are fully diffeomorphism invariant. If the
gravitational theory had no boundary there would be no
diffeomorphism invariant quantities, except for possible
topological invariants. Since we have a boundary, diffeomorphism
invariant quantities are defined at the boundary.

Note that scalar quantities are not invariant. To be invariant they must be scalar and constant.
We therefore need to construct scalar functions that are invariant under changes or radial coordinates.

We can fix this reparametrization invariance by picking a very special frame.
For example choosing the (string) metric in the conformal frame
\be
ds^2=e^{2A}\left[dr^2+dx^{\m}dx_{\mu}\right]\sp \l(r)
\ee
or in the domain-wall frame
\be
ds^2=du^2+e^{2A}dx^{\m}dx_{\m}\sp \l(u)
\ee
fixes the radial reparametrizations almost completely. In conformal frame,
 common scalings of $r,x^{\mu}$ are allowed, corresponding to constant
shifts of $A(r)$.

Eventually we are led to calculate and compare our results to
other ways of calculating (like the lattice).
Some outputs are easier to compare (for example correlators).
Others are much harder as they are not invariant (like the value
 of the coupling at a given energy scale).

In the UV such questions are well understood. The asymptotic energy scale is fixed by comparison to conformal field theory
examples. This is possible because the space is asymptotically AdS$_5$\footnote{As
the dilaton is now not constant there is a non-trivial question: in which frame
is the metric
AdS. In \cite{diss} it was argued that this should be the case in the string
frame. The difference of course between the string and Einstein frame is
 subleading  in the UV as the coupling constant vanishes logarithmically. But this
 may not be the case in the IR where we have very few criteria to check.
 In the model we are using we impose that the space is asymptotically AdS in the
 Einstein frame as this is the only choice consistent with the whole framework.}.

The coupling constant is also fixed to leading order from the coupling
 of the dilaton to $D_3$ branes (up to an overall multiplicative factor).
 Subleading (in perturbation theory) redefinitions of the coupling constant
 and the energy lead to changes in the $\beta$-function beyond two loops.

More in detail, as it  has been described in \cite{ihqcd,diss},
the general form of the kinetic term for the gauge fields on a $D_3$ brane is expected to be:
\be
S_{F^2}=e^{-\phi}Z(R,\xi)Tr[F^2]\sp \xi\equiv -e^{2\phi}{F^2_5\over 5!}
\ee
where $Z(R,\xi)$ is an (unknown) function of curvature $R$ and the five-form field strength, $\xi$.
At weak background fields, $Z\simeq -{1\over 4}+\cdots$.
In the UV regime, expanding near the boundary in powers of the coupling $\lambda \equiv N_c e^{\phi}$ we obtain, \cite{diss}
\be
S_{F^2}=N_c~Tr[F^2]~{1\over \lambda}\left[Z(R_*,\xi_*)-{Z_{\xi}(R_*,\xi_*)\over F_{\xi\xi}(R_*,\xi_*)\sqrt{\xi_*}}
{\l\over \ell}+{\cal O}(\l^2)\right]
\ee
where $F(R,\xi)$ is the bulk effective action and $R_*,\xi_*$ are the boundary values for these parameters.
Therefore the true 't Hooft coupling of QCD is
\be
\lambda_{\rm 't ~Hooft}={\l\over Z(R_*,\xi_*)}\left[1+{Z_{\xi}(R_*,\xi_*)\over Z(R_*,\xi_*)F_{\xi\xi}(R_*,\xi_*)\sqrt{\xi_*}}
{\l\over \ell}+{\cal O}(\l^2)\right] \;.
\ee
In the IR, more important changes can appear between our $\l$ and other definitions as for example in lattice calculations.

In the region of strong coupling we know much less
 in order to be guided concerning the correct definition of the energy.
We can obtain some hints however by comparing with lattice
results.\footnote{We would like to thank K. Kajantie for asking the question, suggesting to compare with lattice
 data, and providing the appropriate references.} In particular, based on lattice calculations using
the Schr\"ondiger functional approach \cite{sommer}, it is argued that at long
distance $L$ the 't Hooft coupling constant scales as \be \l_{\rm
lat}\sim e^{m L}\sp m\simeq {3\over 4}m_{0^{++}} \;. \ee This was
based on a specific definition of the coupling constant, and
length scale on the lattice as well as on numerical data, and some
general expectations on the fall-off of correlations in a massive
theory. This suggests an IR $\beta$ function of the form
\be
L{d\l\over dL}={\l}\log{\l\over \l_0}\sp \l=\l_0 ~e^{mL} \;.
\ee

On the other hand our $\beta$ function at strong coupling uses the UV definition of energy,  $\log E=A_E$
 (the scale factor in the Einstein frame),  $E\sim 1/L$ and is
\be
L{d\l\over dL}={3\over 2}\l\left[1+{3\over 4}{a-1\over
a}{1\over \log\l}+\cdots\right]\sp \l\simeq ~\left({L\over L_0}\right)^{3\over 2} \;.
\label{beta-n}
\ee Consider now taking as length scale the string
scale factor $e^{A_s}$ in the IR. \footnote{The string scale
factor is not a monotonic function on the whole manifold,
\cite{ihqcd} and this is the reason that it was not taken as a
global energy scale. In particular in the UV, $e^{A_s}$ decreases
until it reaches a minimum. The existence of the minimum is
crucial for confinement. After this minimum $e^{A_s}$ increases
and diverges at the IR singularity.}  Since it increases, it
is consistent to consider it as a monotonic function of length.
From its relation to the Einstein scale factor $A_s=A_E+{2\over
3}\log \l$ and (\ref{beta-n}) we obtain \be {d\l\over
dA_s}={2a\over a-1}\l\log\l +\cdots \;.\ee Therefore if we define as
length scale in the IR \be \log L={2a\over
a-1}A_s~~~\to~~~L=\left(e^{A_s}\right)^{2a\over a-1} \ee we obtain
a  running of the coupling compatible with the given lattice
scheme. Note however that $L=\left(e^{A_s}\right)^{2a\over a-1}$
cannot be a global choice but should be only valid in the IR. The
reason is that this function is not globally monotonic.

We conclude this section by restating that physical observables are independent of scheme.
But observables like the 't Hooft coupling constant do depend on schemes, and it is obvious
that our scheme is very different from lattice schemes in the IR.

\section{The potential and the parameters of the model}\label{parameters}

We will make  the following ansatz for  the potential,
\be\label{potential}
V(\l)  = {12\over \ell^2} \left\{ 1 + V_0 \l + V_1 \l^{4/3} \left[\log \left(1 + V_2 \l^{4/3} + V_3 \l^2\right) \right]^{1/2} \right\} ,
\ee
which interpolates between  the two asymptotic
behaviors (\ref{UVexp}) for small $\l$ and (\ref{IRexp}) for large $\l$, with $Q=2/3$ and $P=1/2$. Not
all the parameters entering this potential have physical relevance.
Below we will discuss the independent parameters of the model, and their physical meaning.

\paragraph{The normalization of the coupling constant $\l$.} As discussed  in the previous section,
the relation between the bulk field $\l(r)$ and the physical QCD
't Hooft coupling $\l_t = g_{YM}^2 N_c$ is a priori unknown.
In the UV, the identification of the $D3$-brane coupling to the dilaton implies that
the relation is linear,  and depends on an {\em a priori} unknown coefficient $\kappa$,
defined as:
\be\label{linear}
\l=\kappa\l_t.
\ee
The coefficient $\kappa$ can in principle be
identified by relating the perturbative UV  expansion of the Yang-Mills
$\beta$-function, to the holographic $\beta$-function for the bulk field $\l$:
\bea
&&\beta(\l_t) = -\beta_0 \l_t^2 - \beta_1 \l_t^3 + \ldots \qquad \beta_0 = {22\over 3 (4\pi)^2}, \;\; \beta_1 = {51\over121}\, \beta_0^2 , \; \ldots  \label{beta1}\\
&& \beta(\l) =  -b_0 \l^2 -b_1 \l^3 +\ldots, \qquad b_0 = {9\over 8} v_0, \; \; b_1 = \frac94 v_1 - \frac{207}{256}v_0^2 \,\,\ldots \;.\label{beta2}
\eea

The two expressions (\ref{beta1}) and (\ref{beta2}) are consistent with a linear relation as in (\ref{linear}), and expanding the identity  $\kappa \beta_t(\l_t) = \beta(\kappa \l_t)$  to lowest
order leads to:
\be\label{kappa}
\kappa = \beta_0/b_0.
\ee
Therefore, to relate the bulk field $\l$ to the true coupling $\l_t$ one looks at the linear term in
the expansion of the potential. More generally, the other $\beta$ function coefficients are related
by  $\b_n = \kappa^{n+1} b_n$, and the combinations   $b_n/b_0^{n+1}=\beta_n/\beta_0^{n+1}$ are $\kappa$-independent (however they are scheme-dependent for $n\geq 2$).

As discussed in Section \ref{scheme},
 the introduction of the coefficient $\kappa$ amounts  to a field redefinition and
therefore its precise value does not affect physical (scheme-independent) quantities. In this sense, $\kappa$ is not a parameter that can be fixed by matching some observable computed in the theory.
 Assuming the validity of
the relation (\ref{linear}),  we could  eventually  fix $\kappa$
by matching  a RG-invariant (but scheme-dependent) quantity, e.g.
$\l$ at a given energy scale.

However, as we discuss later in this section, rescaling $\l$ in
the potential (thus changing $\kappa$)
 affects other parameters in the models,
 that are defined in the string frame, e.g. the fundamental string length $\ell_s$: if we hold
the physical QCD string tension fixed, the ratio $(\ell_s/\ell)$
scales with degree $-2/3$ under a rescaling of $\kappa$.

An important point to keep in mind, is that
 the simple linear relation (\ref{linear}) may be modified at strong coupling, but
again this does not have any effect on physical observables.  {\em As long
as we compute RG-invariant and scheme-independent quantities, knowledge of
the exact relationship $\l = F(\l_t)$ is unnecessary. }

\paragraph{The $AdS$ scale $\ell$.} This is set by
the overall normalization of the potential, and its choice  is
equivalent to fixing the unit of energy. It does not enter
dimensionless physical quantities.  As usual the $AdS$ length at
large $N_c$ is much larger than the Planck length ($\ell_p \sim
1/(M_pN_c^{2/3})$, independently of  the 't Hooft coupling.

\paragraph{The UV expansion coefficients of $V(\l)$.} They can be fixed order by order by matching
the Yang-Mills $\b$-function. We impose  this matching up to
two-loops in the perturbative expansion, i.e. $O(\l^3)$ in
$\b(\l)$.  One could go to higher orders  by adding additional
powers of $\l$ inside the logarithm, but since our purpose is not
to give an accurate description of the theory in the UV, we choose
not to introduce extra parameters\footnote{Moreover, higher order
$\beta$-function coefficients are known to be scheme-dependent.}.

Identifying the energy scale with the Einstein frame scale factor, $\log E \equiv \log b(r)$,  we have
the relation (\ref{beta2}) between the $\beta$-function coefficients  and the expansion parameters of $V(\l)$, with
\be
v_0 = V_0, \quad  v_1  = V_1 \sqrt{V_2}.
\ee
The term proportional to $V_2$ in eq. (\ref{potential}) is needed to reproduce  the correct value
of the quantity $b_1/b_0^2 = \b_1/\b_0^2 =51/121$, which is invariant under rescaling of $\l$. Thus, $V_2$ is
not a free parameter, but  is fixed  in terms of $V_0$ and $V_1$ by:
\be\label{V2}
V_2 = b_0^4 \left({23  + 36\, b_1/b_0^2 \over 81 V_1 }\right)^2, \qquad b_0 = {9\over 8} V_0, \quad {b_1 \over b_0^2} = {51\over 121}.
\ee

As explained earlier in this section, when discussing the
normalization of the coupling, fixing the  coefficient $V_0$  is
the same as fixing  the normalization $\kappa$ through eq.
(\ref{kappa}). As we argued,  the actual value of $\kappa$ should
not have any physical consequences, so it is tempting to set
$V_0=1$ by a field redefinition, $\l \to \l/V_0$ and eliminate
this parameter altogether.

In fact, most
of the quantities we will compute are not sensitive to the value of $V_0$, but for certain quantities, such
as the string tension,  some extra care is needed.
In general, we can ask whether two models of the same
form (\ref{a1}),  but  with different  potentials $V(\l)$ and  $\tilde{V}(\l)$, such
that $\tilde{V}(\l) = V(\alpha \l)$ for some constant $\alpha$,  lead to different physical predictions. As
we can change from one model to the other simply by a
 field redefinition $\l \to \alpha \l$ ( this
 has no effect on the other terms in the action in the Einstein frame,
eq. (\ref{a1}) ), clearly the two potentials lead to the same result  for
any physical quantity that can be
 computed unambiguously from the Einstein frame action,
e.g. dimensionless ratios between glueball masses, critical temperature, latent heat etc.

However a rescaling of $\l$ does affect  the string frame metric,
since the latter  explicitly contains factors of $\l$: $b_s(r) = b(r) \l^{2/3}$  \cite{ihqcd} thus,
under the rescaling $\l \to \alpha \l$,   $b_s(r) \to \alpha^{2/3} b_s(r)$.
 This means that any dimensionless ratio of two quantities, such that one of them remains
fixed in the  string frame and  the other in the Einstein frame,
will depend on $\alpha$. An example of this is the ratio $\ell_s/\ell$, where
$\ell_s$ is the string length, that we will discuss shortly.

Therefore, we  can safely perform a field redefinition and set $V_0$ to a given value,
as long as we are careful
when computing quantities that depend explicitly  on the fundamental string length.

Bearing this caveat in mind, we will choose a normalization such that $b_0 = \beta_0$, i.e.
\be\label{V0}
V_0= {8\over 9} \beta_0,
\ee
so that the normalization of $\l$ in the UV matches the physical Yang-Mills coupling. With this choice,
 out of the four free parameters $V_i$ appearing in (\ref{potential})
only $V_1$ and $V_3$ play a non-trivial role ($V_2$ being fixed by eq. (\ref{V2})).

\paragraph{The 5D Planck scale $M_p$.} $M_p$ appears in the overall normalization of the 5D action (\ref{a1}). Therefore it
enters  the overall scale of quantities derived by evaluating the
on-shell action, e.g. the free energy and the black hole mass. It
also sets the conversion factor between the entropy and horizon
area. $M_p$ cannot be fixed directly as we lack a detailed underlining string theory for YM.
 To obtain quantitative predictions, $M_p$  must be fixed in
terms of the other dimension-full quantity of the model, namely the
$AdS$ scale $\ell$. As shown in \cite{GKMN2} this can be done by
imposing that the high-temperature limit of the black hole free
energy be that of a {\em free} gluon gas with the correct number of degrees of freedom\footnote{Note that this is conceptually different from the
${\cal N}=4$
case. There, near the boundary, the theory is strongly coupled and this number must be calculated in string theory.
 It is different by a factor of 3/4 from the free sYM answer. Here near the boundary the theory is free.
Therefore the number of degrees of freedom can be directly inferred.}. This
requires:
\be\label{Planck} (M_p \ell)^3 = {1\over 45\pi^2}. \ee

\paragraph{The string length.} In the non-critical approach the relation between the string length $\ell_s$ and the
5D Planck length (or the $AdS$ length $\ell$)  is not known from first principles. The string length does not appear explicitly in the
2-derivative action (\ref{a1}), but it enters quantities like the static quark-antiquark potential.
The ratio  $\ell_s/\ell$  can be fixed phenomenologically to match the lattice results for the confining
string tension.

More in detail, the relation between the fundamental and the confining string tensions $T_f$ and $\sigma$ is given by:
\be\label{tension}
\sigma  = T_f  \, b^2(r_*) \l^{4/3}(r_*),
\ee
where $r_*$ is the point where the string frame scale factor, $b_s(r) \equiv b(r) \l^{2/3}(r)$, has
its minimum. Fixing the confining string tension by comparison with the lattice result we can
find $T_f$ (more precisely, the dimensionless quantity $T_f \ell^2$, since
the overall scale of the metric depends on $\ell$).
The string length is in turn  given
by  $\ell_s/\ell = 1/\sqrt{2\pi T_f \ell^2}$.

As is clear from eq. (\ref{tension}),  rescaling $\l \to \alpha \l$,  keeping  the value of the QCD string tension $\sigma$  and  of the AdS scale $\ell$ fixed, affects the fundamental string length  in AdS units
as $\ell_s/\ell \to  \alpha^{-2/3}(\ell_s/\ell) $. Therefore two models $a$ and $b$,  {\em defined in the Einstein frame} by eq. (\ref{a1}), but with potentials related by $V_b(\l) = V_a(\alpha \l)$,
must have different fundamental string tensions in order to reproduce the same result for
the QCD string tension. The quantity $\ell_s/\ell$ therefore depends on the value of $V_0$.\\

\paragraph{Integration constants.} Besides the parameters appearing directly
 in the gravitational action, there are also other physically relevant
 quantities that label different solutions to the 5-th order system of field equations.
Any solution is characterized by  a scale $\Lambda$, the
temperature $T$ and a value for the gluon condensate ${\cal G}$,
that correspond to three of the five independent integration
constants.\footnote{The remaining two are the value $f(0)$ which
should be set to one for the solution (\ref{a7})  to obey the right UV
asymptotics, and an unphysical degree of freedom in the
reparametrization of the radial coordinate.}

Regularity at the horizon
fixes  ${\cal G}$ as a function of $T$, so that effectively the gluon condensate
is a temperature-dependent quantity.

The quantity $\Lambda$ controls the asymptotic form of the
solution, as it enters the dilaton running in the UV: $\l \simeq
-(b_0  \log r \Lambda )^{-1}$. It can be defined in a
reparametrization invariant way as: \be \label{Lambda} \Lambda =
\ell^{-1} \lim_{\l\to 0} \left\{b(\l){\exp\left[ - {1\over
b_0\l}\right] \over \l^{b_1/b_0^2}} \right\}, \ee and it is fixed
once we specify the value of the scale factor $b(\l)$ at a given
$\l_0$.

Every choice of $\Lambda$
corresponds to an inequivalent class of solutions, that differ by UV  boundary conditions.
Each class is thermodynamically isolated, since solutions with different $\Lambda$'s have
infinite action difference. Thus, in the canonical partition sum we need to consider
only solutions with  a fixed value of $\Lambda$. However, this choice is merely a choice
of scale, as solutions with different $\Lambda$'s will give the same predictions
for any dimensionless quantity.
In short,  $\Lambda$ is the holographic dual  to the  QCD strong coupling scale: it
is defined by the initial condition to the holographic RG equations, and does not
affect dimensionless quantities such as mass ratios, etc. Therefore, as long
as all solutions we consider obey the same UV asymptotics, the actual value of $\Lambda$ is
immaterial, since the physical units of the system can always be set by fixing $\ell$.\\

To summarize, the only {\em nontrivial} {\it phenomenological} parameters we have at our disposal are $V_1$ and $V_3$ appearing
in (\ref{potential}). The other quantities that enter our model are either fixed by the arguments
presented in this section, or they only affect trivially (e.g. by overall rescaling that can be absorbed in the definition of the
fundamental string scale)
the physical quantities.

In the next section we  present a numerical analysis of the solutions and thermodynamics
of the model defined by eq. (\ref{potential}), and show that for an appropriate choice of the
parameters it reproduces the lattice results for the Yang-Mills deconfinement transition and
high-temperature phase as well as the zero temperature glueball data.

\section{Matching  the thermodynamics of  large-$N_c$ YM} \label{matching}

Assuming a potential of the form (\ref{potential}), we look for
values of the parameters such that the thermodynamics of the 5D
model match the lattice results for the thermodynamics of 4D YM.
As explained in Section \ref{scheme}, we set $V_0$ and $V_2$ as in eqs.
(\ref{V0}) and (\ref{V2}),
 respectively, with $b_0=\beta_0 = 22/3 (4\pi)^{-2}$.

We then vary  $V_1$ and $V_3$ only. We fix these parameters by
looking at thermodynamic quantities corresponding to the latent
heat per unit volume, and the pressure at one value of the temperature above
the transition, which we take as $2T_c$.

It is worth remarking
 that $V_1$ and $V_3$ are phenomenological parameters that we use to fit {\em dimensionless}
QCD quantities. The single (dimension-full) parameter of pure
Yang-Mills, the strong coupling scale, is an extra input that
fixes the overall energy  scale of our solution.

\begin{figure}[h]
\begin{center}
\includegraphics[scale=1.4]{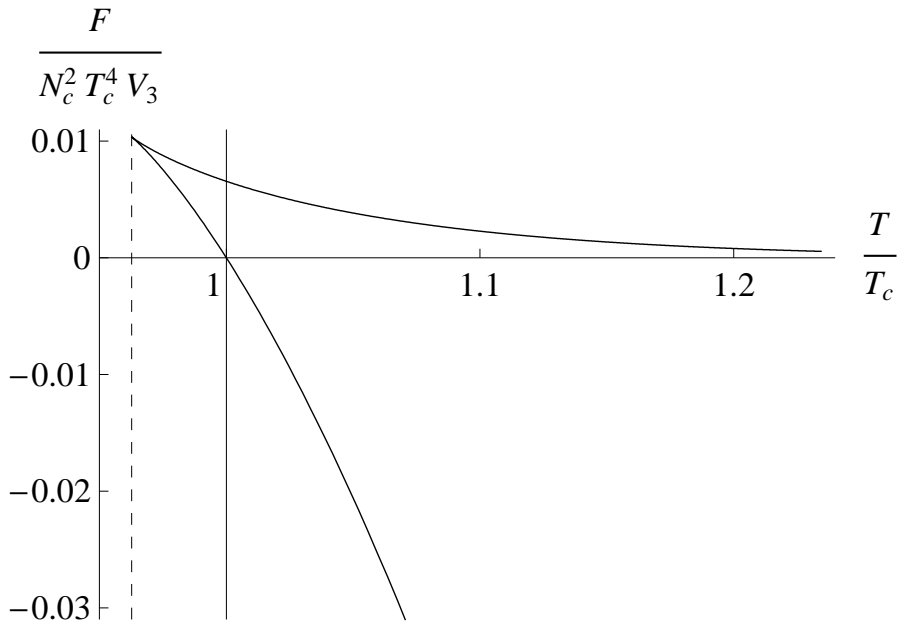}
\end{center}
 \caption[]{The Free energy density (in units of $T_{c}$) as a function of $T/T_c$, for $V_1=14$ and $V_3=170$. The vertical lines correspond to the
critical temperature (solid) and the minimum black hole temperature (dashed). }
\label{FT}
 \end{figure}

\begin{figure}[h!]
 \begin{center}
\includegraphics[height=8cm,width=12cm]{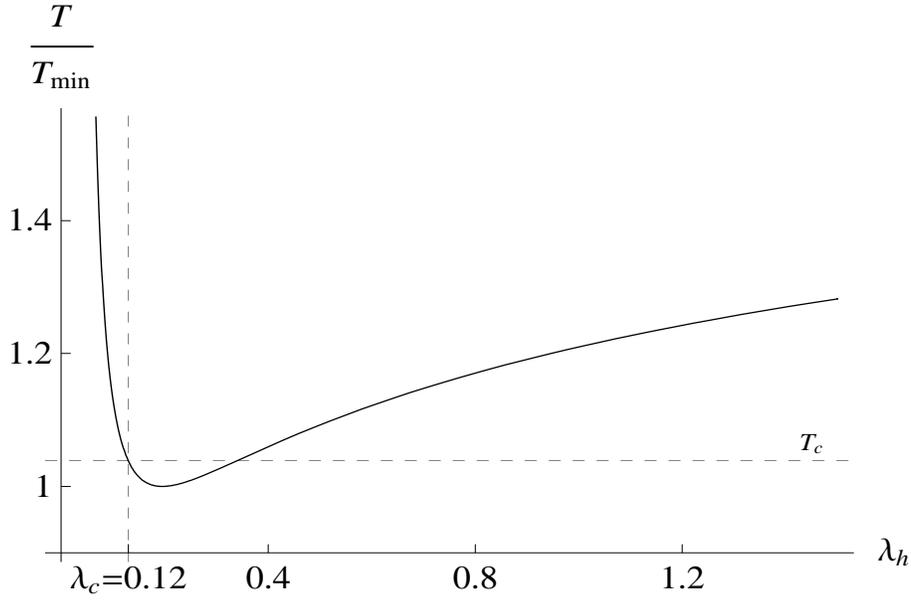}
\end{center}
 \caption[]{Temperature in units of $T_{min}$,  as a function of $\l_h$, for $V_1=14$ and $V_3=170$. The dashed
horizontal and vertical lines indicate the critical temperature and the critical value of the dilaton field at
at the horizon}\label{Tfig}
\end{figure}

\begin{figure}[h!]
 \begin{center}
\includegraphics[height=8cm,width=12cm]{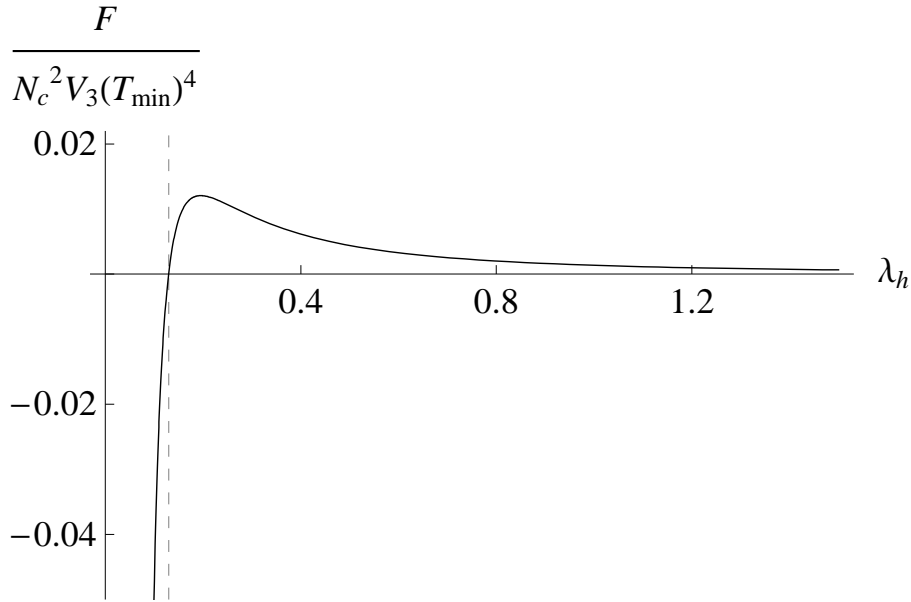}
\end{center}
 \caption[]{The free energy density in units of $T_{min}$,  as a function of $\l_h$}\label{Ffig}
\end{figure}

\begin{figure}[h!]
 \begin{center}
\includegraphics[height=8cm,width=12cm]{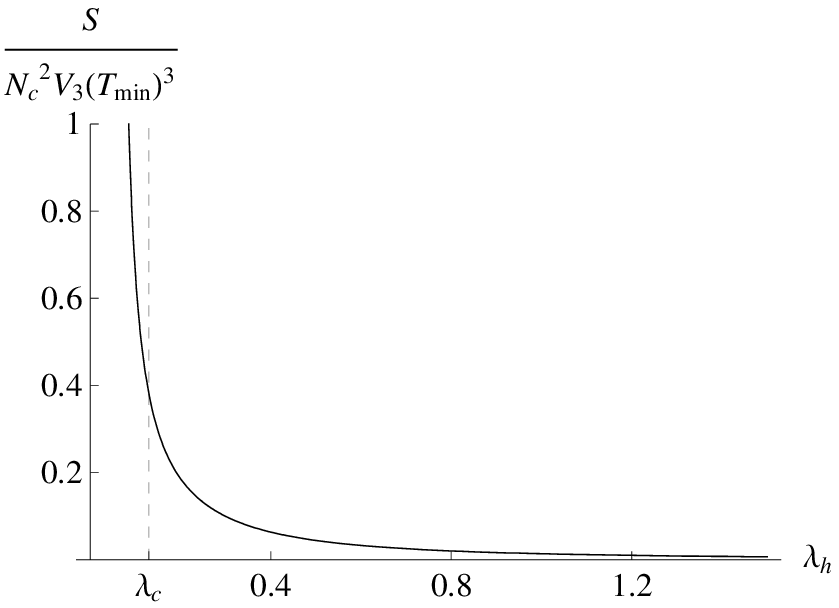}
\end{center}
 \caption[]{Entropy density  in units of $T_{min}$,  as a function of $\l_h$}
\label{Sfig}
\end{figure}

 Using the numerical method explained in the Appendix, for each set of
parameters $(V_1,V_3)$ we numerically generate black hole solutions for  a range of values of $\l_h$,
then from the metric at the horizon and its derivative
 we extract the temperature and entropy functions $T(\l_h)$ and $S(\l_h)$, and the function ${\cal F}(\l_h)$
 from the integrated form of the first law, eq (\ref{intF}).
The behavior of these  functions is  shown in Figures \ref{Tfig}, \ref{Ffig} and \ref{Sfig}, for the best fit
parameter values that we discuss below.
 One  can see the existence of a minimal temperature $T_{min} = T(\l_{min})$,
 and a critical value $\l_c$ where ${\cal F}$ changes sign.
 The resulting function ${\cal F}(T)$
is shown in Figure \ref{FT}.

The phase transition is first order, and the latent heat per unit
volume $L_h$, normalized by $N_c^2 T_c^4$, is given by
the derivative of the curve  in Fig. \ref{FT} at $T/T_c=1$.
Equivalently, $L_h$ is proportional to the jump in the entropy
density $s=S/V_3$ at the phase transition from the thermal gas
(whose entropy is of $O(1)$, in the limit $N_c\to \infty$) to the
black hole (whose entropy scales as $N_c^2$ in the same limit):
thus, in the large $N_c$ limit, \be \label{lh} L_h \equiv T \Delta s
\simeq T_c s(\l_c) \ee
 up to terms of $O(1/N_c^2)$.


At this point we can look for values of $V_1$ and $V_3$ that best fit the lattice data for the deconfined phase
of thermal Yang-Mills. We compare our results  to the data of G. Boyd et al. \cite{karsch}.
 The relevant quantities to compare  are the dimensionless ratios $p(T)/T^4$, $e(T)/T^4$
and $s(T)/T^3$, where $p={\cal F}/V_3$ is the pressure, and $e = p
+ Ts$ is the energy density. Lattice results for these functions
are available in the range $T = T_c \sim 5T_c$, and can be seen in
Figure 7 of \cite{karsch}. The analysis of \cite{karsch}
correspond to  $N_c=3$, but one expects that the thermodynamic
functions do not change to much for large $N_c$\footnote{See e.g.
\cite{lucini}, in which results for $N_c=8$ do not different
significantly from those for $N_c=3$. We thank B. Lucini for
useful correspondence on this point.}.

An additional quantity of relevance is  the value for the ``dimensionless'' latent heat per unit volume, $L_h/T_c^4$
which for large $N_c$ was found in \cite{teperlucini} to be  $(L_h/T_c^4)_{lat} = 0.31 N_c^2$. The result
 for  $N_c^2=3$ is slightly lower ( $\simeq 0.28 N_c^2$).

As already noted in \cite{GKMN1,GKMN2}, the qualitative features of the thermodynamic functions
are generically reproduced in our setup: the curves $3 p(T)/T^4$, $e(T)/T^4$ and $3s(T)/4T^3$ increase
starting at $T_c$, then (very  slowly) approach the  constant free field value $\pi^2 N_c^2/15$ (given
by the Stefan-Boltzmann law) as $T$ increases.
By  computing the thermodynamic functions for various sets of  values of $V_1$ and $V_3$ we find that
1) $V_1$ roughly controls the height reached by the curves $p(T)/T^4$, $e(T)/T^4$ and $s(T)/T^3$ at large $T/T_c$
($\sim$ a few): for larger $V_1$ the curves approach the free field limit faster;  2) $V_3$ does not
affect much the height of the curves at large $T/T_c$, but on the other hand it changes the latent heat, which is increasing as $V_3$
decreases.

We find that the best fit corresponds to the values

\be\label{bestfit}
V_1= 14 \qquad
V_3 = 170.
\ee
Below we discuss the values of  various physical  quantities ( both related to thermodynamics, and to
zero-temperature properties) obtained with this choice of parameters.

\subsection{Latent heat and equation of state}

\begin{figure}[h]
 \begin{center}
\includegraphics[scale=1.4]{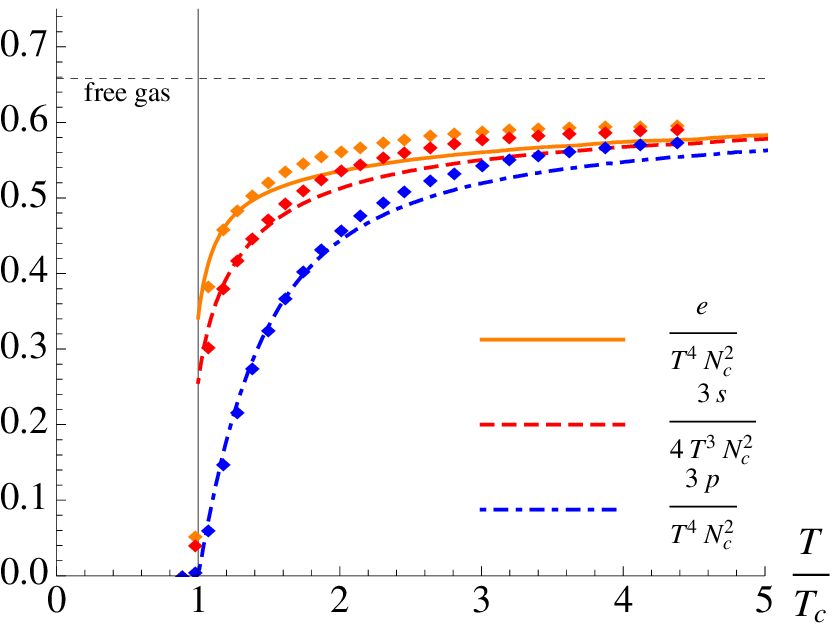}

\end{center}
 \caption[]{Temperature  dependence of the dimensionless thermodynamic densities
 $s/T^3$ (light blue), $p/T^4$ (dark blue) and $e/T^4$ (green),
  normalized such that they reach the common limiting value $\pi^2 /15$ (dashed
horizontal line) as $T\to \infty$. The dots correspond to the lattice data for $N_c=3$ \cite{karsch}. }
\label{esp}
\end{figure}

The comparison between the curves $p(T)/T^4$, $e(T)/T^4$ and
$s(T)/T^3$
 obtained in our models with (\ref{bestfit}),   and the lattice results  \cite{karsch}\footnote{We thank
F. Karsch for providing us the relevant data.} is shown in Figure \ref{esp}.
The match is remarkably good
for $T_c<T<2T_c$, and deviates slightly  from the lattice data in the range up to $5T_c$.

The latent heat we obtain is:
\be
 L_h/T_c^4 = 0.31 N_c^2,
\ee
which matches the lattice result for $N_c \to \infty$ \cite{teperlucini}.

\begin{figure}[h]
 \begin{center}
\includegraphics[scale=1.4]{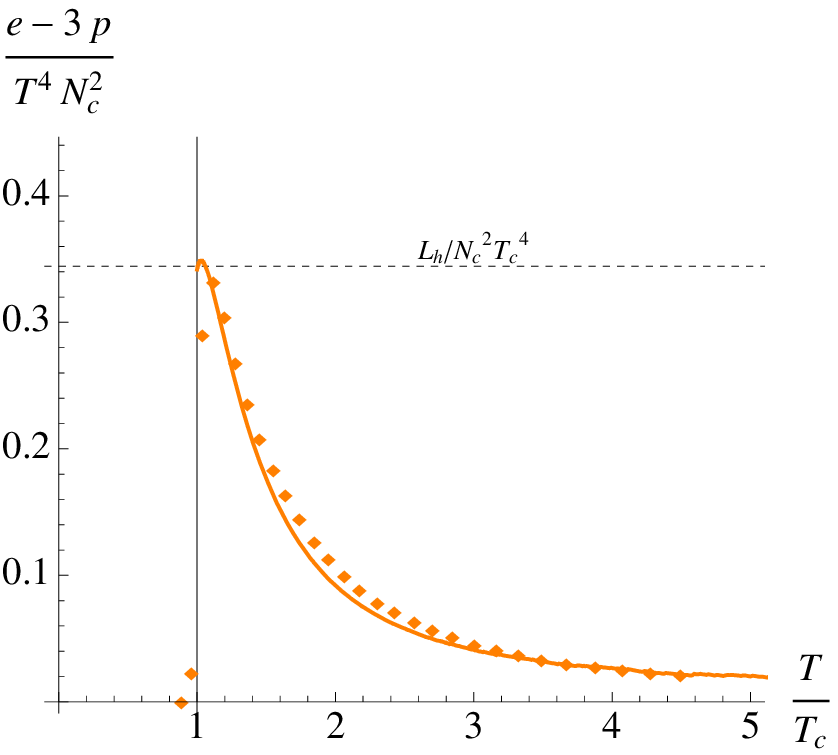}
\end{center}
 \caption[]{The trace anomaly as a function of temperature  in the deconfined phase
of the holographic model  (solid line) and the corresponding
lattice data \cite{karsch} for $N_c=3$ (dots). The peak in the lattice data slightly above $T_c$ is expected to be
an artifact of the finite lattice volume. In the infinite volume limit the maximum value
of the curve  is at $T_c$, and it equals $L_h/N_c^2 T_c^4$.}
\label{trace}
\end{figure}

An interesting quantity is the {\em trace anomaly} $(e-3p)/T^4$,
(also known as {\em interaction measure}), that indicates the
deviation from conformality, and it is proportional to the gluon
condensate. The trace anomaly in our setup is shown, together with
the corresponding lattice data, in figure (\ref{trace}), and the
agreement is again very good.

\begin{figure}[h]
 \begin{center}
\includegraphics[scale=1.4]{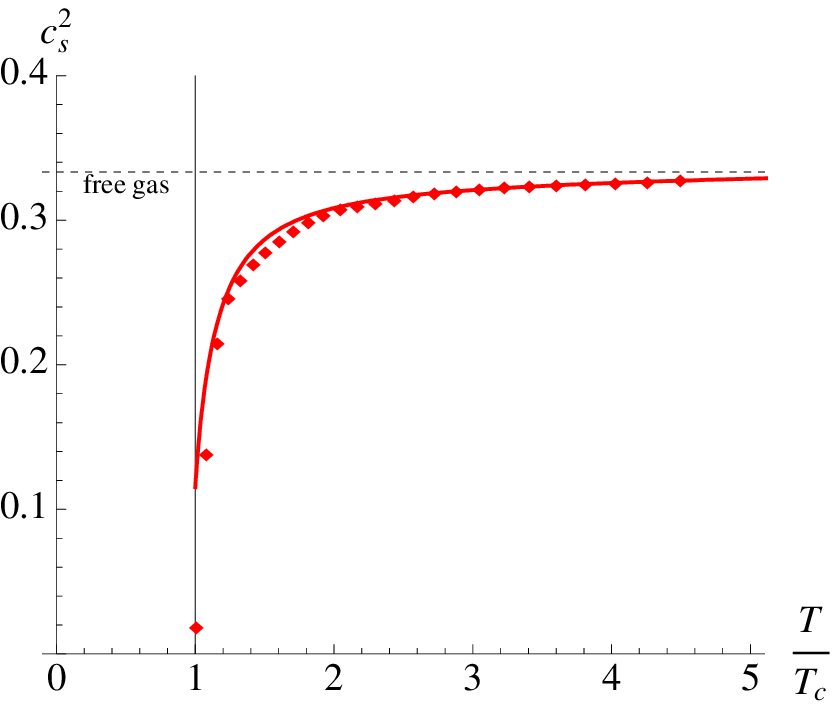}
\end{center}
 \caption[]{The speed of sound in the deconfined phase, as a function of temperature, for the
holographic model (solid line) and the corresponding   lattice data \cite{karsch} for $N_c=3$ (dots).
The dashed horizontal line indicates the conformal limit $c_s^2 = 1/3$.}
\label{sound}
\end{figure}

\begin{figure}[h]
 \begin{center}
\includegraphics[scale=1.4]{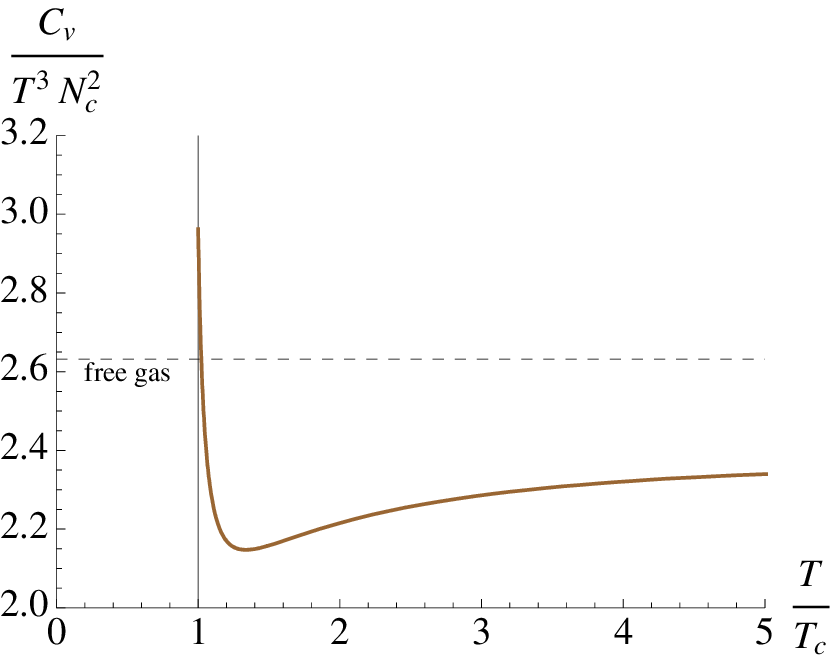}
\end{center}
 \caption[]{The specific heat (divided by $T^3$\textmd{}), as a function of temperature, in the
deconfined phase of the holographic model.}
\label{cv}
\end{figure}

We also compute the specific heat per unit volume $c_v$, and the
speed of sound $c_s$ in the deconfined phase,  by the  relations
\be\label{cvcs} c_v = - T {\de^2 F \over \de T^2}, \qquad \quad
c_s^2 = {s\over c_v} \;. \ee These are shown in Figures \ref{cv} and
\ref{sound} respectively. The speed of sound is shown together
with the lattice data, and the agreement is remarkable.

\subsection{Glueball spectrum}

In \cite{ihqcd}, the single phenomenological parameters of the potential was fixed
by looking at the zero-temperature spectrum,
 i.e. by computing various glueball mass ratios and comparing
them  to the corresponding lattice results.
The associated thermodynamics for this potential was studied in \cite{GKMN1}
which was in qualitative agreement with lattice QCD results, but not in full quantitative agreement.
This is due to the fact that the thermodynamics depends more on the details of the potential than the glueball spectrum for the main Regge trajectories.
Here we use  the potential (\ref{potential}), but with the two phenomenological parameters
$V_1$ and $V_3$ already determined by the thermodynamics
(\ref{bestfit}).

The glueball spectrum is obtained holographically as the spectrum
of normalizable fluctuations around the zero-temperature
background.
As explained in the introduction, and motivated in \cite{ihqcd,diss}, here we consider
explicitly  the 5D  metric, one scalar field
(the dilaton), and one pseudoscalar field (the axion).
 As a consequence, the only normalizable
fluctuations above the vacuum correspond to spin 0 and spin 2 glueballs\footnote{Spin 1 excitations of the
metric can be shown to be non-normalizable.} (more precisely, states
 with $J^{PC} = 0^{++}, 0^{-+}, 2^{++}$), each species containing an infinite discrete tower of excited
states.

In 4D YM there are many more operators generating glueballs, corresponding to different values of $J^{PC}$,
that are not considered here. These are expected to correspond holographically to other fields
in the noncritical string spectrum (e.g. form fields, which may yield  spin 1 and CP-odd spin 2 states) 
and to higher string states that provide higher spin glueballs. As the main focus of this paper
is reproducing  the YM thermodynamics in detail rather than the entire glueball spectrum, we chose not to include
these states\footnote{A futher  reason is that, unlike the scalar and (to some extent)
the pseudoscalar sector that we are considering, the action governing the higher Regge slopes 
is less and less universal as one goes to higher masses. Only a precise knowledge of the underline string 
theory is expected to provide detailed information for such states.}. Therefore we only compare the  mass spectrum
obtained in our model to the  lattice results for the lowest $0^{++}, 0^{-+}, 2^{++}$ glueballs and their
available excited states. These are limited to one for each spin 0 species, and none for the spin 2,
 in the study of \cite{chenetal}, which is the one we use for our  comparison.
This provides  two mass ratios in the CP-even sector and two  in the CP-odd sector.

The glueball masses are computed by first  solving numerically
the zero-temperature  Einstein's equations, obtained from
(\ref{numeqs}) by setting $f(r)=1$, and using the resulting metric
and dilaton to setup an  analogous  Schr\"odinger problem for the
fluctuations, \cite{ihqcd}.
The results for the parity-conserving sector are shown in Table
\ref{masses++}, and are in good agreement with those reported by
\cite{chenetal} for $N_c=3$, whereas the
results reported by \cite{teperlucini2} for large $N_c$ are somewhat larger.
 The CP-violating  sector (axial glueballs)  will be
discussed separately.

\TABLE[h!]{
\begin{tabular}{|r|c|c|c|}
\hline
& HQCD & $N_c=3$ \cite{chenetal} & $N_c=\infty$ \cite{teperlucini2} \\
\hline\hline
$m_{0^{*++}}/m_{0^{++}}$ & 1.61 & 1.56(11) & 1.90(17)  \\
\hline
$m_{2^{++}}/m_{0^{++}}$ & 1.36 & 1.40(4) & 1.46(11)   \\
\hline \hline
\end{tabular}
\caption{Glueball Masses}\label{masses++}
}

We should add that there are other lattice studies (see e.g. \cite{meyer}) that report
 additional excited states. Our mass ratios offer a somewhat worse fit of the mass ratios found in  \cite{meyer} (whose results are
not entirely compatible with those of \cite{chenetal} for the states the two studies have in common).
We should stress however that  reproducing the detailed glueball spectrum  is secondary here since  the main focus is  thermodynamics. 
However, the comparison of our spectrum to the existing lattice results shows that our model provides a good global fit to 4D YM also with respect to
quantities beyond thermodynamics.

Unlike the various  mass ratios,
the value of  any given mass in $AdS$-length units (e.g. $m_{0++} \ell$)  {\em does  depend}
 on the choice of integration constants in the UV, i.e.
on the value of $b_{UV}$ and $\l_{UV}$. Therefore its numerical
value does not have an intrinsic meaning. However it can be used
as a benchmark against which all other dimension-full quantities can
be measured (provided one always uses the same UV b.c. ).  On the
other hand, given a fixed set of initial conditions, asking that
$m_{0++}$ matches the physical value (in MeV) obtained on the
lattice, fixes the value of $\ell$ hence the energy unit.

\subsection{Critical Temperature}

The thermodynamic quantities we have discussed so far, are
dimensionless ratios, in units of the critical temperature. To
compute $T_c$,   we need an extra dimension-full quantity which can
be used independently to set the unit of energy. In lattice
studies this is typically the confining string tension $\sigma$ in
the $T=0$ vacuum, with a value of around $(440 MeV)^2$, and
results are given in terms of the dimensionless ratio
$T_c/\sqrt{\sigma}$. In our case we cannot compute $\sigma$
directly, since it depends on the {\em fundamental} string
tension, which is a priori unknown. Instead, we take the mass
$m_{0++}$ of the lowest-lying glueball state as a reference.

We compute  $m_{0++}$ with  the potential  (\ref{potential}), with
$V_1$ and $V_3$ fixed as in (\ref{bestfit}),  then compare
$T_c/m_{0++}$ to the same quantity obtained on the lattice. For
the  lattice result, we  take the large $N_c$ result of
\cite{teperlucini},  $T_c/\sqrt{\sigma} = 0.5970(38)$, and combine
it with the large $N_c$ result for the lowest-lying glueball  mass
\cite{teperlucini2},  $m_{0++}/\sqrt{\sigma}= 3.37(15)$. The two
results are in fair  agreement, without need to adjust any extra
parameter: \be \left(T_c\over m_0\right)_{hQCD}  = 0.167, \qquad
\left(T_c\over m_0\right)_{lattice}  = 0.177(7) \;.\ee In physical
units, the critical temperature we obtain is given by \be T_c =
0.56\, \sqrt{\sigma}  =  247 \, MeV. \ee

\subsection{String tension}

 The fundamental string tension $T_{f}={1\over 2\pi\ell_s^2}$ cannot be computed from first principles in our model, but can be
obtained using as extra input the lattice value of the confining
string tension $\sigma$, at $T=0$. The fundamental and
confining string tensions are related by eq. (\ref{tension}).
As
for the critical temperature, we can relate $T_f$ to the 
to the value of the lowest-lying glueball mass,
by using the lattice relation  $\sqrt{\sigma} = m_{0++}/3.37  $ \cite{teperlucini2}. Since what
we actually compute numerically is $m_{0++} \ell$, this allows us to obtain the string tension $T_f$ (and fundamental
string length $\ell_s = 1/\sqrt{2\pi T_f}$) in $AdS$ units. 
Evaluating  numerically the factor $b(r_*)\l^{2/3}(r_*)$ that appears in eq. (\ref{tension}), we find: 
\be\label{tension-numbers}
T_f  \ell^2 = 6.5,  \qquad \ell_s/\ell = 0.15 \;.
\ee
This shows that the fundamental string length in our model, although not 
parametrically small, turns out to be one order of magnitude smaller than  the  $AdS$ length, which is a good sign for the validity of the derivative expansion. 
The meaning of this fact is a little  more complicated conceptually, as
the discussion in \cite{diss} indicates. Also, we should
stress that, as discussed in Section \ref{parameters}, this
result depends on our choice of the overall normalization of $\l$:
changing the potential
 by $\l \to \kappa \l$ will yield different
 numerical values in (\ref{tension-numbers}) without affecting
the other physical quantities.

\subsection{CP-odd sector}

The CP-odd sector of pure Yang-Mills is described holographically
by the addition of a bulk  pseudoscalar field $a(r)$ (the {\em
axion}) with action:\footnote{This action was justified in \cite{ihqcd,diss}. The dilaton dependent coefficient $Z(\l)$ is
encoding both the dilaton dependence as well as the UV curvature dependence of the axion kinetic terms in the associated string theory.
We cannot determine it directly from the string theory,
but we pin it down by a combination of first principles and lattice input, as we explain further below.}
\be\label{axionaction} S_{axion} =
{M_p^3\over2} \int d^5 x  Z(\l) \sqrt{-g} (\de^\mu a) (\de_\mu a) \;.
\ee
The field $a(r)$ is dual to the topological density operator
$\tr F\tilde{F}$. The prefactor $Z(\l)$ is  a dilaton-dependent
normalization. The axion action is suppressed by a factor
$1/N_c^2$ with respect to the action (\ref{a1}) for the dilaton and the
metric, meaning that in the large-$N_c$ limit one can neglect
the back-reaction of the axion on the background.

As shown in \cite{ihqcd}, requiring the correct scaling  of $a(r)$
in the UV, and phenomenologically  consistent axial glueball
masses, constrain the asymptotics of $Z(\l)$ as follows:
\be\label{Z1} Z(\l)  \sim Z_0 \;,\; \l \to 0; \qquad Z(\l) \sim \l^4
\;,\;\l \to \infty, \ee where $Z_0$ is a constant.
 As a simple interpolating function between
these large- and small- $\l$ asymptotics we can take the
following: \be\label{Z2} Z(\l)  = Z_0(1 + c_a \l^4). \ee The
parameter $Z_0$ can be fixed by matching the topological
susceptibility of pure Yang-Mills theory,
 whereas $c_a$ can be fixed by looking at the axial glueball mass spectrum.

\paragraph{Axial glueballs.}
As in \cite{ihqcd}, we can fix $c_a$ by matching to the
lattice results the mass ratio $m_{0-+}/m_{0++}$ between the lowest-lying axial and scalar glueball states.
This is independent of the overall coefficient $Z_0$ in (\ref{Z2}). The lattice value $m_{0-+}/m_{0++} = 1.49$ \cite{chenetal}
is obtained for:
\be\label{ca}
c_a = 0.26.
\ee
  With this choice, the mass of the first excited axial glueball  state
is in good agreement  with the corresponding lattice result
\cite{chenetal}: \be \left({m_{0-+*}\over m_{0++}}\right)_{hQCD} =
2.10 \qquad \left({m_{0-+*}\over m_{0++}}\right)_{lattice} =
2.12(10) \;.\ee

\paragraph{Topological Susceptibility.}
In pure Yang-Mills, the topological $\chi$  susceptibility is
defined by: \be\label{chi1} E(\theta) = {1\over 2}\, \chi \,
\theta^2, \ee where $E(\theta)$ is the vacuum energy density in
presence of a $\theta$-parameter. $E(\theta)$ can be computed
holographically by solving for the  axion profile $a(r)$ on a
given background, and evaluating the action  (\ref{axionaction})
on-shell.

In the deconfined phase, the axion profile is trivial, implying a vanishing
topological susceptibility \cite{GKMN2}. This is in agreement
with large-$N_c$ arguments and lattice results \cite{ltheta}.

\begin{figure}[h]
 \begin{center}
\includegraphics[scale=1.4]{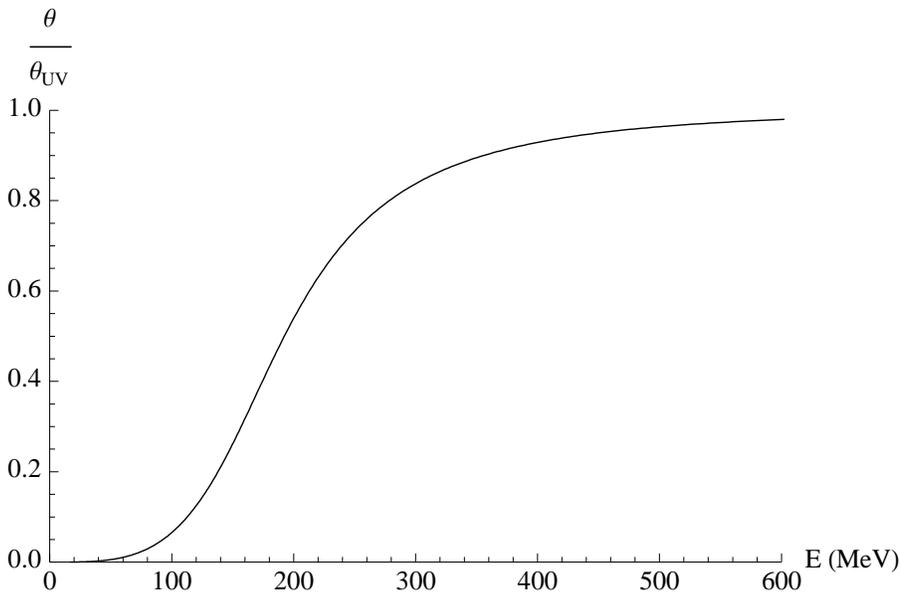}
\end{center}
 \caption[]{Axion profile in the radial direction. The  $x$-axis is
taken to be the energy scale, $E(r)=E_0 b(r)$, where the unit $E_0$ is
fixed to match the lowest glueball mass.}
\label{axionprofile}
\end{figure}

In the low-temperature phase, the axion acquires a non-trivial profile,
\be\label{aprofile}
a(r) = a_{UV} \, {F(r)\over F(0)}, \qquad F(r)\equiv \int_r^{\infty} {dr \over Z(\l(r)) e^{3A(r)} } \;.
\ee
This profile is shown, for the case at hand, in Figure \ref{axionprofile}, where
the axion is normalized to its UV value.

The topological susceptibility is given by \cite{ihqcd}:
\be\label{chi2} \chi = M_p^3 F(0)^{-1} = M_p^3\left[ \int_0^\infty
{dr \over e^{3A(r)}Z(r)} \right]^{-1}, \ee where $Z(r)\equiv
Z(\l(r))$. Evaluating this expression numerically with $Z(\l)$ as
in (\ref{Z2}),  and $c_a=0.26$ (to match the axial glueball
spectrum ),  we can determine the coefficient $Z_0$ by looking at
the lattice result for $\chi$. For $N_c=3$,  \cite{DelDebbio}
obtained $\chi = (191 MeV)^4$, which requires $Z_0 = 133$.


In Table \ref{parameters} we present a summary of the various
physical quantities discussed in this section, as obtained in our
holographic  model, and their comparison with the lattice results
for large $N_c$ (when available) and for $N_c=3$. The quantities
shown in the upper half of the table  are the ones that were used
to fix the free parameters (reported in the last column)  of the
holographic model.

\subsection{Coupling normalization}

Finally, we can relate the field $\l(r)$ to the running 't Hooft
coupling. All other quantities  we have discussed so far are
scheme-independent and RG-invariant. This is not the case for the
identification of the physical YM 't Hooft coupling,  which is
scheme dependent.

\begin{figure}[h]
 \begin{center}
\includegraphics[scale=1.4]{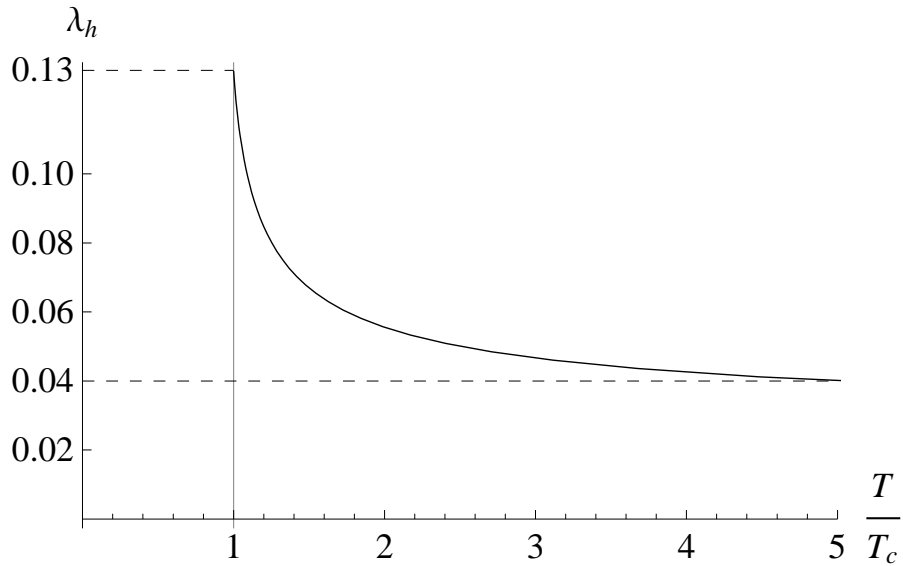}
\end{center}
 \caption[]{The coupling at the horizon as a function of  temperature in the range $T_c$--$5T_c$.}
\label{couplingvsT}
\end{figure}

In the black hole phase we can take $\l_h \equiv \l(r_h)$ as a measure of the
temperature-dependent coupling. In figure \ref{couplingvsT} we show $\l_h$ as a function of the temperature
in the range $T_c$ to  $5T_c$.

As a reference, we may take the result of \cite{karsch}, that
found $g^2(5T_c)\simeq 1.5$ for $N_c=3$, which translates to
$\l_t(5T_c) \simeq 5$. On the other hand,  if we  make the
assumption that the  identification $\l = \l_t$  is valid at all
scales (not only in the UV), we find in our model $\l_t(5T_c)
\simeq 0.04$ (see Figure \ref{couplingvsT}), i.e. a factor of 100
smaller than the lattice result.

 This discrepancy is almost certainly due to the identification
(\ref{linear})being very different from lattice at strong coupling.

\TABLE[h]{
\begin{tabular}{|r|c|c|c|l|}
\hline
& HQCD &  lattice $N_c=3$  & lattice $N_c\to \infty$ & Parameter \\
\hline \hline
&&&&\\
$[p/(N_c^2 T^4)]_{T=2T_c}$ & {\bf 1.2} & {\bf 1.2} & -  & $V1=14$\\
&&&&\\
$L_h/(N_c^2 T_c^4)$ & {\bf 0.31} & 0.28 \cite{karsch} & {\bf 0.31} \cite{teperlucini} & $V3=170$ \\
&&&&\\
$[p/(N_c^2 T^4)]_{T\to +\infty}$ & {\bf $\pi^2/45$} & {\bf $\pi^2/45$}  &{\bf $\pi^2/45$ }& $M_{p} \ell=[45 \pi^2]^{-1/3}$ \\
&&&&\\
$m_{0^{++}}/\sqrt{\sigma}$ & {\bf 3.37}  & 3.56 \cite{chenetal} &{\bf  3.37} \cite{teperlucini2}&$\ell_s/\ell =0.15$\\
&&&&\\
$m_{0^{-+}}/m_{0^{++}}$ & {\bf 1.49} &  {\bf 1.49} \cite{chenetal}  &- & $c_a= 0.26 $\\
&&&&\\
$\chi$ & {\bf $(191 MeV)^4$} &  $ (191 MeV)^4$ \cite{DelDebbio} & - & $Z_0=133$ \\
&&&&\\
\hline\hline
&&&&\\
$T_c/m_{0^{++}}$ & 0.167 &- &0.177(7) &  \\
&&&&\\
\hline
&&&&\\
$m_{0^{*++}}/m_{0^{++}}$ &1.61 &1.56(11) &1.90(17) & \\
&&&&\\
$m_{2^{++}}/m_{0^{++}}$ & 1.36 & 1.40(4)& 1.46(11)& \\
&&&&\\
\hline
&&&&\\
$m_{0^{*-+}}/m_{0^{++}}$ & 2.10 & 2.12(10) & - & \\
&&&&\\
\hline
\hline
\end{tabular}
\caption{Collected in this table is the complete set of physical quantities that we computed in our model and compared with data.
The upper half of the table contains the quantities that we used as input (shown in boldface) for the holographic QCD model (HQCD). Each quantity  can be roughly associated to one parameter of the model (last column). The lower half of the table
contains our ``postdictions'' (i.e. quantities that we computed after all the parameters were fixed) and the
comparison with the corresponding lattice results.  The value we find for  the critical temperature corresponds to $T_c=247 MeV$.
}\label{results}
}

\section{Discussion and Outlook}\label{conclusions}

The construction presented in this paper offers a holographic description of large-$N_c$ Yang-Mills theory that
is both realistic and calculable, and in quite good agreement with a large number of lattice results
both at zero and finite temperature.

It is a phenomenological model; as such it is not directly associated to an
explicit string theory construction.
In this respect it is  in the same class as the models
based  on pure $AdS$ backgrounds (with hard or soft walls) \cite{adsqcd1,soft,herzog,braga}, or on IR deformations of the $AdS$ metric \cite{adr,kaj}.
In comparison to them however, our approach  has the
advantage that the dynamics responsible for strong coupling phenomena (such as
confinement and phase transitions)  is made  explicit in the bulk description,
and  it is tied  to  the fact that the coupling constant depends on the energy scale and becomes
large in the IR. This makes the model consistent and calculable,
once the 5D effective action is specified: the dynamics can be entirely derived from the bulk Einstein's equation.
The emergence of an IR mass scale and the finite temperature phase structure are built-in:
they  need not be imposed by hand and
do not suffer from ambiguities related to IR boundary conditions (as in hard wall models) or from inconsistencies
in the laws of thermodynamics (as in non-dynamical soft wall models based on a fixed dilaton profile \cite{soft} or on a fixed metric \cite{adr,kaj}).
More specifically,  in our approach it is guaranteed that
the Bekenstein-Hawking temperature of the black hole matches the entropy computed as the derivative of the free energy
with respect to temperature.

With an appropriate choice of the potential,  we provide a realistic and quantitatively accurate description of
essentially all the static properties (spectrum and equilibrium thermodynamics) of the dynamics of pure Yang-Mills.
The main ingredient responsible for the
dynamics (the dilaton potential) is fixed through comparison  with both perturbative QCD and
lattice results. It is worth stressing that  such a matching on the quantitative level was  only possible
because the class of holographic models we discuss {\em generically} provides a {\em qualitatively accurate}
description of the strong Yang-Mills dynamics. This is a highly  non-trivial fact, that strongly indicates
that a realistic  holographic description of real-world QCD might be ultimately possible.

Although the asymptotics of our potential is dictated by general principles, we based our choice of parameters
by comparing with the lattice results for the thermodynamics.
There are other physical parameters in the 5D description  that do not appear in the potential:
the 5D Planck scale, that was fixed by matching the free field thermodynamics
in the limit  $T\to \infty$; the coefficients in the axion kinetic term, that were set by matching
the axial glueball spectrum and the topological susceptibility (from the lattice).  The quantities that
we use as input in our fit, as well as the corresponding parameters in the 5D model, are shown
in the upper half of Table \ref{results}.

The fact that our  potential has effectively two free parameters depends on our choice of the functional form. This
functional form contains some degree of arbitrariness, in that only
 the UV and IR asymptotics of $V(\l)$  are fixed by general considerations
(matching the perturbative $\beta$-function in the UV, and  a discrete linear glueball spectrum for the IR).
Therefore the results presented in this paper offer more a {\em description}, rather than a {\em prediction} of the
thermodynamics.

Nevertheless, there are several quantities that we successfully ``postdict'' (i.e. they agree with
the lattice results) once the potential is fixed: apart from the good agreement of the thermodynamic functions
over the whole range of temperature explored by the lattice studies (see Figures \ref{esp}, \ref{trace} and \ref{sound}),
they are the lowest glueball mass ratios and the value
of the critical temperature.  The comparison  of these quantities with the lattice results is shown in the lower half
of Table \ref{results}, and one can see that the agreement is overall very good.
Moreover the model predicts the masses of the full towers of glueball states in the $0^{+-},0^{++},2^{++}$ families.

The fact that our model is consistent with a large number of lattice results is clearly
not the end of the story: its added value, and one of the main reasons for its interest lies in
its immediate applicability beyond equilibrium thermodynamics, i.e. in the dynamic regimes tested
in heavy-ion collision experiments. This is a generic feature of the holographic approach, in which
there are no obstructions (as opposed to the lattice) to perform real-time computations and to calculate
hydrodynamics and transport coefficients.
Our model provides a framework to compute these quantities
in a case where the static properties agree with the real-word  QCD at the quantitative level, and this will be reported in a future publication.

In the approach followed in this work, we took the phenomenological two-derivative bulk action at face value and
applied the standard  rules of holography to relate it to  a 4D theory. Once the 5D action is taken as an input,
this procedure is well defined, and yields unique and well defined results. The smallness of the gravitational coupling
 in the large $N_c$ limit ensures that quantum corrections can be neglected, independently of the details of the
underlying fundamental theory.  Here we did not discuss the justification  of
this approach, and its possible limits of validity,  in the broader context  of non-critical string theory. These
are still open issues, that were addressed in \cite{ihqcd} and more  recently in  \cite{diss}.

\section*{Acknowledgments}
We would like to thank O. Aharony, R. Emparan , B. Fiol, S. Gubser,  K. Kajantie, F. Karsch, D. Kharzeev, B. Lucini, D. Mateos, I. Papadimitriou, D. T. Son,
  for useful discussions. This work was  supported by the VIDI grant 016.069.313 from the Dutch Organization
for Scientific Research (NWO), ANR grant NT05-1-41861 and CNRS PICS  3747, INFN.

Elias Kiritsis is on leave of absence from CPHT, Ecole Polytechnique (UMR du CNRS 7644).

\newpage
\appendix
\renewcommand{\theequation}{\thesection.\arabic{equation}}
\addcontentsline{toc}{section}{Appendix}

\section*{APPENDIX}

\section{Numerical technique}\label{numerics}

To compute the thermodynamics from the 5D holographic model we need to find different temperature  black hole solutions of Einstein-Dilaton field  equations with the potential (\ref{potential}). This  cannot be done analytically, and
we must resort to numerical integration. The main difficulty in reaching this goal is how to fix the integration
constant, or equivalently boundary conditions
so that the following requirements are met:
\begin{itemize}
\item[i.] all the solutions must obey the same UV asymptotics as $r\to 0$. In particular $f(r)$ should go to one a the the boundary and
the integration constant $\Lambda$ discussed in the Section \ref{parameters} should be the same for all solutions.
\item[ii.] the solutions must have regular horizons, rather than naked singularities. As shown in \cite{GKMN2},
this imposes one constrain among  the values of $b(r)$, $\dot{b}(r)$, $\dot{f}(r)$, and $\l(r)$ evaluated at the
``would be'' horizon, i.e. where $f = 0$.
\end{itemize}
Clearly the first requirement privileges a numerical integration starting from the UV, while the second one starting
from the horizon. One elegant way out of this dilemma is to use  a solution-generating technique based on a set of symmetry
transformations   enjoyed by the field equations: we starts numerical integration at the horizon imposing regular  initial conditions,
obtaining thus a regular solution; then we act with
the symmetry transformations in order to generate a new solution that satisfy given UV boundary conditions. The second step can
be done analytically, and since the symmetry transformation does not spoil regularity of the horizon, the new solution
can satisfy both the above requirements.

The equations of motion can be written in the following form
\bea \label{numeqs}
&\dot \l(r) = \frac{3}{2} \l(r) \sqrt{b(r) \dot W(r)} \quad,\quad \dot b(r) = - b(r) W(r) \quad, &\\
&\ddot f(r) = 3 b(r) \dot f(r) W(r) \quad,\quad \dot W(r) = 4 b(r) W(r)^2 - \frac{\dot f(r)}{f(r)} W(r) - \frac{b(r)}{3f(r)} V(\l(r)) \quad,& \non
\eea
where $W(r)$ is an auxiliary function that we introduce for convenience and reduces to the zero-temperature superpotential in the thermal-gas solution \cite{GKMN2}.

The five integration constants discussed in the previous Section are fixed in the numerical procedure by giving initial conditions at the horizon and
by exploiting the symmetries of the system of equations \refeq{numeqs} to rescale the solutions at different temperatures so that they describe the
same theory, i.e. with the same UV behavior (or, in other words, with the same\footnote{The solutions at different temperature will also be
characterized by the same normalization for $f$ in the UV and the same location of the UV boundary, $r_{UV}$.} $\Lambda$). More precisely, we set the
five initial conditions for $\l(r)$, $b(r)$, $f(r)$, $\dot f(r)$, $W(r)$ at a point $r=r_i$ close to the horizon using the expansion of these functions to the
first order in $\e_h \equiv (r_h-r_i)$ (the dots in the formulae stand for ${\cal O}\left((r_h-r_i)^2\right)$ terms):
\bea
&b(r_i) = b_h + \dot b_h \e_h + \dots \quad , \quad
\l(r_i) = \l_h + \dot \l_h \e_h + \dots \quad , & \label{h1}\\
&f(r_i) = \dot{f}_h \e_h + \dots \quad,\quad W(r_i) = W_h + \dot
W_h \e_h + \dots \quad, &\non \eea where we have introduced the
notation $\dot{b}_h \equiv \dot{b}(r_h)$ and similarly for the
other dotted quantities appearing in the equations above.

The solution is uniquely determined by specifying the five
quantities  $\{\l_h$, $b_h$, $W_h$, $\dot{f}_h$, $\e_h\}$ at the position
$r_h$\footnote{The ``true'' horizon position of the physical
solution is not $r_h$, but its image under the solution generating
transformations which we will introduce below. Therefore the value
$r_h$ is immaterial and can be chosen arbitrarily.},  since the
linear (for small $\e_h$)  system (\ref{h1}) can be translated to
an initial value  problem at $r_i$ with full set of five initial
data, $\{\l(r_i)$, $b(r_i)$, $W(r_i)$, $f(r_i)$, $\dot{f}(r_i)\}$.

Regularity at the horizon constrains  the initial data to obey the relation \cite{GKMN2}:
\be\label{regularity}
W_h=\frac{b_h
V(\l_h)}{3\dot{f}_h} \;.
\ee
 The other first order coefficients $\dot b_h,\dot W_h,\dot \l_h$ are all determined by solving the
equations of motion close to the horizon.

Among the (free) parameters characterizing the initial conditions
$\l_h, b_h, \dot{f}_h, \e_h$, we vary the physical parameter
$\l_h$ alone to scan solutions with different horizon value of
$\lambda(r)$, while $b_h, f_h, \e_h$ are kept fixed (and can be
chosen arbitrarily). For each $\l_h$ we generate numerically  a
solution $b(r),\l(r),f(r)$. Apart for the value of $\l(r_h)$,
these these solutions all share the same near-horizon behavior,
 but for the time being  they all have different UV asymptotics (including different values of $f(0)$).

The second step of the numerical procedure is to match the solutions in the UV. In fact, in order for the solutions obtained
for  different $\l_h$'s  to describe the same theory, we have to make sure that we have the same asymptotic behavior for the scale
factor and the dilaton (i.e. the same $\Lambda$), and  the same normalization for $f$. For convenience (but
this  is not a strict physical requirement), we can
also ask that all the solutions have coinciding UV boundary, rather than coinciding horizon. For each solution obtained numerically,
 we can  generate a new solution obeying these three requirements
by acting  with the following  three transformations (parametrized by the constants $\delta_f$, $\delta_b$ and $\delta_r$):

\begin{enumerate}
\item scaling of $b$ and $f$:\\
$ b(r) \to b(r)/\delta_f $,
$ \l(r) \to \l(r) $,
$ f(r) \to f(r)/\delta_f^2 $,
$ W(r) \to W(r) \delta_f $;  
\item scaling of $b$ and $r$:\\
$ b(r) \to b(r\delta_b)/\delta_b$,
$ \l(r) \to \l(r\delta_b)$,
$ f(r) \to f(r\delta_b) $,
$ W(r) \to W(r\delta_b) $;
\item shift in $r$:\\
$b(r) \to b(r-\delta_r) $,
$ \l(r) \to \l(r-\delta_r) $,
$ f(r) \to f(r-\delta_r)$,
$ W(r) \to W(r-\delta_r) $.
\end{enumerate}

These transformations  leave the system of equations \refeq{numeqs} invariant, and they preserve the regularity condition
(\ref{regularity}). Thus, they map a regular solution into another regular solution. Moreover, they do not act on the
function $\l(r)$, in particular they leave $\l_h=\l(r_h)$ invariant (although they change the other horizon quantities,
including the horizon position $r_h$).

The
first  transformation can be  used to set the right normalization of $f(r)$ in the UV, $f(r_{UV})=1$.
Acting with the second  transformation changes the UV behavior of the metric and dilaton, or equivalently $\Lambda$ (without spoiling the UV value of $f$); hence the strong coupling scale can be
fixed by using this symmetry to obtain the same UV asymptotic behavior for all solutions, yielding the same value $\Lambda$ from eq.\refeq{Lambda}.
Finally, the shift in $r$ simply serves as a mechanism to fix the same
UV boundary for all solutions.

Explicitly, for each $\l_h$,  given the solution $\{\l(r),b(r),f(r),W(r)\}$,
 acting with the three transformations above we generate a new solution  $\{\tilde \l(\tilde r),\tilde b(\tilde r),\tilde f(\tilde r),\tilde W(\tilde r)\}$:
\bea\label{tildesol}
&\tilde b(\tilde r) = b(r)/(\delta_b \delta_f) \quad,\quad \tilde \l(\tilde r) = \l(r) \quad, &\\
& \tilde f(\tilde r) = f(r)/\delta_f^2 \quad,\quad \tilde W(\tilde r) = W(r)\delta_f \quad, & \non
\eea
where $\tilde r = r\delta_b - \delta_r$.  For each $\l_h$, the parameters $\delta_b(\l_h), \delta_f(\l_h), \delta_r(\l_h)$ are  determined such that
the UV behavior is the same for all $\l_h$.  Namely: the UV boundary is at $\tilde r=\tilde r_{UV}$, the scale factor and dilaton match close to the
UV boundary $\tilde b(\tilde r_{UV}+\tilde \e)=\tilde b_{UV}$, $\tilde \l(\tilde r_{UV}+\tilde \e)=\tilde \l_{UV}$ (with given constants $\tilde r_{UV}$, $\tilde
b_{UV}$, $\tilde \l_{UV}$ that can be chosen arbitrarily\footnote{Setting $\tilde b_{UV}$ and $\tilde \l_{UV}$ determines the energy scale of the
theory, namely $\Lambda\ell$, while $\tilde r_{UV}$ is irrelevant and can be set to zero for convenience.} and taken to be the same for all solutions). We
first set $\delta_f(\l_h)$ such that
\bea\label{deltaf}
\delta_f(\l_h) = \sqrt{ f(r_{UV}) }\quad,
\eea
then $\delta_b(\l_h)$
\bea\label{deltab}
\delta_b(\l_h) = b ( r_{UV} + \e ) / (\tilde b_{UV} \delta_f(\l_h))
\eea
and finally $\delta_r(\l_h)$
\bea\label{deltar}
\delta_r(\l_h) = r_{UV} \delta_b(\l_h) - \tilde r_{UV} \quad.
\eea
In eq.\refeq{deltab} $\e\ll|r_h - r_{UV}|$ is chosen such that $\l(r_{UV} + \e ) = \tilde \l_{UV}$ for all $\l_h$, implying $\tilde \l(\tilde r_{UV} +
\tilde \e) = \tilde \l_{UV}$ with $\tilde \e = \e \delta_b \ll |\tilde r_h - \tilde r_{UV}|$. The dependence of the r.h.s of
eqs.\refeq{deltaf}--\refeq{deltar} on $\l_h$ is implicit in the dependence of the solutions $b(r)=b(r;\l_h)$, $\l(r)=\l(r;\l_h)$, $f(r)=f(r;\l_h)$  on $\l_h$

The solutions \refeq{tildesol} with the definitions \refeq{deltaf}--\refeq{deltar} now represent the physical solutions that can be used to compute
the thermodynamic properties of the holographic plasma, since they all share the same UV behavior:
\bea
&\tilde r(r_{UV}) = \tilde r_{UV} \quad,\quad \tilde f(\tilde r_{UV}) = 1 \quad, &\\
&\tilde b(\tilde r_{UV} + \tilde \e) = \tilde b_{UV} \quad,\quad \tilde \l(\tilde r_{UV} + \tilde \e) = \tilde \l_{UV} \quad,& \non
\eea
for all different values of $\l_h$. \\

Next, we describe how to compute the thermodynamic properties of a given solution.
\paragraph{The temperature} Inserting  into eq.\refeq{TS} the expression for $\dot f(r_h) = - \dot{f}_h \delta_f^{-2}$ (from now on we drop the $\tilde{}$ to simplify notation),
that is obtained after acting with the aforementioned transformations on the solutions, we get

\bea
T(\l_h) = \frac{f_h}{4\pi \delta_f^2(\l_h)} \quad.
\eea

\paragraph{The free energy}
The free energy is the sum of the two contributions coming from $\cal G$ and $C$ respectively. The entropy contribution $C$ (see eq.\refeq{CG}) is
given by the area of the black-hole horizon $C(\l_h) = \left(b_h/\delta_b(\l_h)\delta_f(\l_h)\right)^3$.

The contribution related to the gluon condensate can be evaluated in two different ways: by directly fitting the metric (or dilaton) fluctuation
$b_o(r) - b(r)$ close to the UV boundary, or by making use of the first law of thermodynamics in the form $S=-\pa {\cal F}/\pa T$. The latter method
is more efficient\footnote{A numerical fit of the fluctuation can be performed in order to compute $\cal G$ following eq.\refeq{b-bo}. It gives larger
errors than the integral method described below, though.}. In particular, summing the two contributions in eq.\refeq{F} yields the following expression
for the free energy

\be\label{intF}
{\cal F}(\l_h) = \int_{\l_h}^{+\infty}\intd\bar\l_h\,  S(\bar \l_h) {\intd T(\bar \l_h) \over \intd\bar \l_h} \quad.
\ee

Here $S(\l_h)$ is given by \refeq{TS} and both the big black-hole
and small black-hole branches are needed in order the get the full
result for the free energy. This is because the integral in
\refeq{intF} extends to $+\infty$, entering deeply in the small
black-hole branch. In the numeric evaluation the integral is cut
off at a large value $\Lambda_h \gg \l_{h,min}$, where ${\cal
F}(\Lambda_h) \ll {\cal F}(\lambda_{h,min})$, with $\l_{h,min}$
corresponding to the minimum temperature at which the two
black-hole branches join. Thus, the value $\Lambda_h$ provides a
systematic way of controlling the numerical error one makes in
using the formula \refeq{intF}.

\paragraph{Thermodynamic functions} All thermodynamic functions are given in terms of the free energy, entropy and temperature. The energy density
$e=E/V_3$ can be read from eq. \refeq{first law}, while the
pressure $p$ follows from $p=-{\cal F}/V_3$.

The speed of sound is given by $c_s^2 = \pa p / \pa e = s / C_v$ (with $C_v$ denoting the specific heat at constant volume).


\subsection{An alternative  technique}

Here we present another numerical method that is used to check the
findings obtained by the previous method above. The main advantage
of this method is the simplicity. In particular, the idea now is
to solve the Einstein's equations using the reduced system of
variables described in Section 7 of \cite{GKMN2}. As explained there,
 the information relevant for the physical
observables at finite temperature is entirely encoded in a coupled
system of two first order differential equations for the scalar
variables $X\equiv \f'/3A'$ and $Y\equiv f'/4f A'$. Note that
these functions are invariant under the diffeomorphisms of the
radial coordinate. The idea is to use $\f$ as the radial
coordinate and solve for $X$ and $Y$ as functions of $\f$. This is
enough to construct thermodynamical observables.

Given the potential $V(\l)$ one first integrates the reduced
Einstein system, \bea\label{Xeq}
\l \frac{dX}{d\l} &=& -\frac43(1-X^2+Y)\le(1+\frac{3}{8X}\frac{\l d\log V}{d\l}\ri),\\
\l \frac{dY}{d\l} &=& -\frac{4}{3}(1-X^2+Y)\frac{Y}{X},
\label{Yeq} \eea from a boundary $\l_0$ (chosen to be arbitrarily
small) to the horizon $\l_h$. In solving (\ref{Xeq}) and
(\ref{Yeq}) one imposes the boundary conditions at the horizon:
\bea\label{XYh} Y&\to& \frac{Y_h}{\log(\l_h/\l)} + {\cal O}(1),\nonumber\\
X&\to& -\frac43 Y_h + {\cal O}(\log(\l_h/\l)), \eea as $\l\to\l_h$
where $Y_h$ is given by $Y_h = 9\l_h V'(\l_h)/32V(\l_h)$.

Now, the thermodynamic observables can be determined by either of
the three methods below:
\begin{enumerate}
\item Similar to (\ref{Xeq}), we solve the same equation that
corresponds to zero T that is obtained by setting $Y=0$ in
(\ref{Xeq}). Let us call the solution of this $X_0$. One has to
specify the boundary condition for numerical integration: Pick an
arbitrary point $\l_*$. Specify the value of $X_0$ at this point
$X(\l_*)=X_*$ such that for $\l\gg\l_*$ $X_0$ asymptotes to the
desired value as required from color confinement \cite{ihqcd}.
For the potential (\ref{potential}) this is $X_0\to -1/2$ for
$\l\gg\l_*$.

Having obtained $Y$, $X$ and $X_0$ one determines the constants of
motion $\Y0$ and $\C0$ by fitting the numerical solution to
following asymptotic forms: \bea\lab{Y0}
Y(\l) &=& \Y0\,\, e^{-\frac{4}{b_0\l}}(b_0\l)^{-4b}, \\
\lab{C0} X(\l) - X_0(\l) &=& \le[\frac{\Y0/2-\C0}{X_0}+\C0
X_0\ri]e^{-\frac{4}{b_0\l}}(b_0\l)^{-4b}, \eea for $\l$ close to
the boundary $\l_0$. The constants of motion $\C0$ and $\Y0$ are
functions of the horizon location $\l_h$ (hence temperature) and
have the physical meaning of energy and enthalpy respectively. To
compute them in terms of $T$ one uses the relation \cite{GKMN2},
\be\lab{Tder3} T = \frac{\Y0}{\pi b^3(\l_h)}
[\ell^{-1}e^{4A_0-\frac{4}{b_0\l_0}}(b_0\l_0)^{-4b}]. \ee Here
$A_0$ is the initial value of the scale factor and can be set to
an arbitrary value as it drops out when we compute thermodynamic
observables as functions of the ratios $T/T_c$. Similarly $\ell$
can be set to 1.

In summary, one first solves for $X$, $Y$ and $X_0$, then
determines the constants of motion $\Y0$ and $\C0$ from (\ref{Y0})
and (\ref{C0}) as functions of $\l_h$, next one translates them
into functions of $T$ using (\ref{Tder3}) and finally obtains the
free energy density from them using \cite{GKMN2}:
\begin{equation}\label{freeCY}
\frac{\cal F(T)}{V_3 N_c^2} =  \frac{\Lambda^4}{45\pi^2}
\le(6\C0(T) - 4\Y0(T)\ri).
\end{equation}
Here $\Lambda$ is the integration constant (\ref{Lambda}) which
should be set by the zero temperature solution, since we require
that the black-hole and the zero T solution have the same
asymptotics near the boundary. Dependence on $\Lambda$ drops out
when we compute dimensionless quantities such as $s/T^3$. $T_c$ is
obtained by the value of $T$ that makes (\ref{freeCY}) zero. Given
(\ref{freeCY}) all the rest of the thermodynamics follow from the
formulae presented in section 7 of \cite{GKMN2}.

\item There is a second method that is simpler. Here, one does not
need to determine $X_0$. One solves for $X$ and $Y$ as above, than
uses the first law of thermodynamics to construct ${\cal F}$
directly from the entropy $S$, using (\ref{intF}). Here $S$ in
(\ref{intF}) is given by the Bekenstein-Hawking formula (\ref{TS})
where the scale factor is constructed from $X$ by \cite{GKMN2}:
\be\lab{Adet} A(\l) =  A_0 + \int_{\l_0}^{\l}
\frac{1}{3X}\frac{d\tilde{\l}}{\tilde{\l}}, \ee whereas the $T$ in
(\ref{intF}) is given by (\ref{Tder3}) above. $T_c$ is determined
by the value that makes (\ref{intF}) zero.

\item We end this subsection by presenting an even simpler
formula, that side-steps most of the complications above and
determines the thermodynamic functions semi-directly, in terms of
the potential. Given $V$, one solves for $X$ as above. Then the
entropy density $s=S/V_3 N_c^2$ is given by \be\lab{sT3}
\frac{s}{T^3} = \frac{768\pi^2}{5\ell^6}
\frac{e^{-4\int_0^{\l_h}\frac{d\l}{\l} X(\l)}}{V(\l_h)^3} =
\frac{16384\pi^2}{1215\ell^3} \le(\frac{W(\l_h)}{V(\l_h)}\ri)^3.
\ee In the second line we used the definition of the thermal
superpotential in terms of $X$, \cite{GKMN2}. We refer to
\cite{GKMN2} for a derivation of (\ref{sT3}). Using this and
(\ref{Tder3}) one easily obtains the thermodynamic functions. We
note that the dimensionless quantity $s/T^3$ is one of the main
functions that is compared with the lattice data, hence this
formula provides a semi-direct method of relating the data to be
compared with lattice and the dilaton potential. The reason for
its being semi-direct rather than direct is because one still has
to determine $T_c$ in order to set the scale which requires use of
(\ref{sT3}) in (\ref{intF}).

\end{enumerate}



\addcontentsline{toc}{section}{References}

\begin{thebibliography}{99}

\bibitem{rhic}
  J.~Adams {\it et al.}  [STAR Collaboration],
  {\em ``Experimental and theoretical challenges in the search for the quark  gluon
  plasma: The STAR collaboration's critical assessment of the  evidence from
  RHIC collisions,''}
  Nucl.\ Phys.\  A {\bf 757} (2005) 102
  \hre{nucl-ex}{0501009};\\
  B.~B.~Back {\it et al.},
  {\em ``The PHOBOS perspective on discoveries at RHIC,''}
  Nucl.\ Phys.\  A {\bf 757} (2005) 28
  \hre{nucl-ex}{0410022};\\
  I.~Arsene {\it et al.}  [BRAHMS Collaboration],
  {\em ``Quark gluon plasma and color glass condensate at RHIC? The perspective
  from the BRAHMS experiment,''}
  Nucl.\ Phys.\  A {\bf 757} (2005) 1
  \hre{nucl-ex}{0410020};\\
   K.~Adcox {\it et al.}  [PHENIX Collaboration],
  {\em ``Formation of dense partonic matter in relativistic nucleus nucleus
  collisions at RHIC: Experimental evaluation by the PHENIX  collaboration,''}
  Nucl.\ Phys.\  A {\bf 757} (2005) 184
  \hre{nucl-ex}{0410003}.

\bibitem{lr}
  M.~Luzum and P.~Romatschke,
  {\em ``Conformal Relativistic Viscous Hydrodynamics:  Applications to RHIC results
  at $\sqrt{s_{NN}}$ = 200 GeV,''}
  Phys.\ Rev.\  C {\bf 78} (2008) 034915
  \hri{0804.4015}{[nucl-th]}.

 \bibitem{pss}
  G.~Policastro, D.~T.~Son and A.~O.~Starinets,
  {\em ``The shear viscosity of strongly coupled N = 4 supersymmetric Yang-Mills
  plasma,''}
  Phys.\ Rev.\ Lett.\  {\bf 87} (2001) 081601
  \hre{hep-th}{0104066};\\
  P.~Kovtun, D.~T.~Son and A.~O.~Starinets,
  {\em ``Viscosity in strongly interacting quantum field theories from black-hole
  physics,''}
  Phys.\ Rev.\ Lett.\  {\bf 94} (2005) 111601
  [arXiv:hep-th/0405231].

\bibitem{review}
  E.~Shuryak,
  {\em ``Physics of Strongly coupled Quark-Gluon Plasma,''}
  \hri{0807.3033}{[hep-ph]};\\
  D.~T.~Son and A.~O.~Starinets,
  {\em ``Viscosity, Black Holes, and Quantum Field Theory,''}
  Ann.\ Rev.\ Nucl.\ Part.\ Sci.\  {\bf 57} (2007) 95
  \hri{0704.0240}{[hep-th]]};\\
  M.~Natsuume,
  {\em ``String theory and quark-gluon plasma,''}
  \hre{hep-ph}{0701201}.
\bibitem{kkt}
  F.~Karsch, D.~Kharzeev and K.~Tuchin,
  {\em ``Universal properties of bulk viscosity near the QCD phase transition,''}
  Phys.\ Lett.\  B {\bf 663} (2008) 217
  \hri{0711.0914}{[hep-ph]}.
\bibitem{m}
  H.~B.~Meyer,
  {\em ``A calculation of the bulk viscosity in SU(3) gluodynamics,''}
  Phys.\ Rev.\ Lett.\  {\bf 100} (2008) 162001
  \hri{0710.3717}{[hep-lat]};
 {\em ``Energy-momentum tensor correlators and viscosity,''}
  \hri{0809.5202}{[hep-lat]}.

\bibitem{adsqcd1}
  J.~Erlich, E.~Katz, D.~T.~Son and M.~A.~Stephanov,
 {\em ``QCD and a holographic model of hadrons,''}
  Phys.\ Rev.\ Lett.\  {\bf 95}, 261602 (2005)
  [arXiv:hep-ph/0501128];\\
L.~Da Rold and A.~Pomarol,
 {\em ``Chiral symmetry breaking from five dimensional spaces,''}
  Nucl.\ Phys.\  B {\bf 721}, 79 (2005)
  \hre{hep-ph}{0501218}.

 \bibitem{soft}
  A.~Karch, E.~Katz, D.~T.~Son and M.~A.~Stephanov,
  {\em ``Linear confinement and AdS/QCD,''}
  Phys.\ Rev.\  D {\bf 74} (2006) 015005
  \hre{hep-ph}{0602229}.

 \bibitem{her}
  C.~P.~Herzog, A.~Karch, P.~Kovtun, C.~Kozcaz and L.~G.~Yaffe,
  {\em ``Energy loss of a heavy quark moving through N = 4 supersymmetric
  Yang-Mills plasma,''}
  JHEP {\bf 0607} (2006) 013;
  \hre{hep-th}{0605158}.

\bibitem{lrw}
  H.~Liu, K.~Rajagopal and U.~A.~Wiedemann,
  {\em ``Calculating the jet quenching parameter from AdS/CFT,''}
  Phys.\ Rev.\ Lett.\  {\bf 97} (2006) 182301
  \hre{hep-ph}{0605178}.

\bibitem{gub}
  S.~S.~Gubser,
  {\em ``Drag force in AdS/CFT,''}
  Phys.\ Rev.\  D {\bf 74} (2006) 126005
  \hre{hep-th}{0605182};\\
  {\em ``Comparing the drag force on heavy quarks in N = 4 super-Yang-Mills theory
  and QCD,''}
  Phys.\ Rev.\  D {\bf 76} (2007) 126003
 \hre{hep-th}{0611272}.

   \bibitem{tea}
  J.~Casalderrey-Solana and D.~Teaney,
  {\em ``Heavy quark diffusion in strongly coupled N = 4 Yang Mills,''}
  Phys.\ Rev.\  D {\bf 74} (2006) 085012
\hre{hep-ph}{0605199}.

\bibitem{ps}
 J.~Polchinski and M.~J.~Strassler,
 {\em ``Hard scattering and gauge/string duality,''}
  Phys.\ Rev.\ Lett.\  {\bf 88} (2002) 031601
  \hre{hep-th}{0109174}.

\bibitem{ihqcd}
  U.~Gursoy and E.~Kiritsis,
 {\em  ``Exploring improved holographic theories for QCD: Part I,''}
  JHEP {\bf 0802} (2008) 032
  \hri{0707.1324 }{[hep-th]};\\
   U.~Gursoy, E.~Kiritsis and F.~Nitti,
  {\em ``Exploring improved holographic theories for QCD: Part II,''}
  JHEP {\bf 0802} (2008) 019
  \hri{0707.1349}{ [hep-th]}.

\bibitem{diss}
  E.~Kiritsis,
  {\em ``Dissecting the string theory dual of QCD,''}
  \hri{0901.1772}{[hep-th]}.

\bibitem{ckp}
  R.~Casero, E.~Kiritsis and A.~Paredes,
  {\em ``Chiral symmetry breaking as open string tachyon condensation,''}
  \hre{hep-th}{0702155}.

\bibitem{GKMN1}
U.~Gursoy, E.~Kiritsis, L.~Mazzanti and F.~Nitti,
  {\em ``Deconfinement and Gluon Plasma Dynamics in Improved Holographic QCD,}
  Phys.\ Rev.\ Lett.\  {\bf 101}, 181601 (2008)
  \hri{0804.0899}{[hep-th]}.


\bibitem{GKMN2}
U.~Gursoy, E.~Kiritsis, L.~Mazzanti and F.~Nitti,
  {\em ``Holography and Thermodynamics of 5D Dilaton-gravity,}
  \hri{0812.0792 }{[hep-th]}.

\bibitem{gubser}
S.~S.~Gubser and A.~Nellore,
 {\em  ``Mimicking the QCD equation of state with a dual black hole,''}
  \hri{0804.0434}{ [hep-th]}.

  \bibitem{dew}
  O.~DeWolfe and C.~Rosen,
  {\em ``Robustness of Sound Speed and Jet Quenching for Gauge/Gravity Models of Hot
  QCD,''}
  \hri{0903.1458}{[hep-th]}.

\bibitem{ltheta}
  E.~Vicari and H.~Panagopoulos,
  {\em ``Theta dependence of SU(N) gauge theories in the presence of a topological
  term,''}
\hri{0803.1593}{[hep-th]}.


\bibitem{D4} E.~Witten,
  {\em ``Anti-de Sitter space, thermal phase transition, and confinement in  gauge  theories,''}
  Adv.\ Theor.\ Math.\ Phys.\  {\bf 2} (1998) 505
  \hre{hep-th}{9803131};



\bibitem{sonnenschein}
Y.~Kinar, E.~Schreiber and J.~Sonnenschein,
  {\em ``Q anti-Q potential from strings in curved spacetime: Classical results,''}
  Nucl.\ Phys.\  B {\bf 566}, 103 (2000)
  \hre{hep-th}{9811192}.

 \bibitem{sommer}
  J.~Heitger, H.~Simma, R.~Sommer and U.~Wolff  [ALPHA collaboration],
  {\em ``The Schroedinger functional coupling in quenched QCD at low energies,''}
  Nucl.\ Phys.\ Proc.\ Suppl.\  {\bf 106} (2002) 859
  \hre{hep-lat}{0110201}.


\bibitem{karsch}
G.~Boyd, J.~Engels, F.~Karsch, E.~Laermann, C.~Legeland, M.~Lutgemeier and B.~Petersson,
  {\em ``Thermodynamics of SU(3) Lattice Gauge Theory,''}
  Nucl.\ Phys.\  B {\bf 469}, 419 (1996)
  \hre{hep-lat}{9602007}.


\bibitem{lucini}
B.~Bringoltz and M.~Teper,
  {\em ``The pressure of the SU(N) lattice gauge theory at large-N,''}
  Phys.\ Lett.\  B {\bf 628}, 113 (2005)
  \hre{hep-lat}{0506034}.



\bibitem{teperlucini}
B.~Lucini, M.~Teper and U.~Wenger,
  {\em ``Properties of the deconfining phase transition in SU(N) gauge theories,''}
 JHEP {\bf 0502}, 033 (2005)
  \hre{hep-lat}{0502003}.


\bibitem{chenetal}
C.~J.~Morningstar and M.~J.~Peardon,
  {\em ``The glueball spectrum from an anisotropic lattice study,''}
  Phys.\ Rev.\  D {\bf 60}, 034509 (1999)
  \hre{hep-lat}{9901004};\\
 Y.~Chen {\it et al.},
  {\em ``Glueball spectrum and matrix elements on anisotropic lattices,''}
  Phys.\ Rev.\  D {\bf 73}, 014516 (2006)
  \hre{hep-lat}{0510074}.


\bibitem{teperlucini2}B.~Lucini and M.~Teper,
 {\em  ``SU(N) gauge theories in four dimensions: Exploring the approach to N =$\infty$},''
  JHEP {\bf 0106}, 050 (2001)
  \hre{hep-lat}{0103027}.

\bibitem{meyer}
H.~B.~Meyer,
  {\em ``Glueball Regge trajectories,''}
  \hre{hep-lat}{0508002}.

\bibitem{DelDebbio}
  L.~Del Debbio, L.~Giusti and C.~Pica,
  Phys.\ Rev.\ Lett.\  {\bf 94}, 032003 (2005)
  \hre{hep-th}{0407052}.

\bibitem{herzog}
  C.~P.~Herzog,
 {\em  ``A holographic prediction of the deconfinement temperature,''}
  Phys.\ Rev.\ Lett.\  {\bf 98}, 091601 (2007)
  \hre{hep-th}{0608151}.
\bibitem{braga}
C.~A.~Ballon Bayona, H.~Boschi-Filho, N.~R.~F.~Braga and L.~A.~Pando Zayas,
  Phys.\ Rev.\  D {\bf 77}, 046002 (2008)
  \hri{0705.1529}{[hep-th]}.

 \bibitem{adr}
  O.~Andreev,
  {\em ``Some Thermodynamic Aspects of Pure Glue, Fuzzy Bags and Gauge/String
  Duality,''}
  Phys.\ Rev.\  D {\bf 76} (2007) 087702
  \hri{0706.3120}{ [hep-ph]}.

\bibitem{kaj}
 K.~Kajantie, T.~Tahkokallio and J.~T.~Yee,
  {\em ``Thermodynamics of AdS/QCD,''}
  JHEP {\bf 0701} (2007) 019
  \hre{hep-ph}{0609254}.







\end{thebibliography}
\end{document}